\documentclass[12pt]{article}
\usepackage{epsfig,epsf}
\begin{document}
\def\beq{\begin{equation}}
\def\eeq{\end{equation}}
\def\eq#1{{Eq.~(\ref{#1})}}
\def\fig#1{{Fig.~\ref{#1}}}
\newcommand{\bas}{\bar{\alpha}_S}
\newcommand{\as}{\alpha_S} 
\newcommand{\bra}[1]{\langle #1 |}
\newcommand{\ket}[1]{|#1\rangle}
\newcommand{\bracket}[2]{\langle #1|#2\rangle}
\newcommand{\intp}[1]{\int \frac{d^4 #1}{(2\pi)^4}}
\newcommand{\mn}{{\mu\nu}}
\newcommand{\tr}{{\rm tr}}
\newcommand{\Tr}{{\rm Tr}}
\newcommand{\T} {\mbox{T}}
\newcommand{\braket}[2]{\langle #1|#2\rangle}
\newcommand{\ab}{\bar{\alpha}_S}

\setcounter{secnumdepth}{7}
\setcounter{tocdepth}{7}
\parskip=\itemsep               

\setlength{\itemsep}{0pt}       
\setlength{\partopsep}{0pt}     
\setlength{\topsep}{0pt}        
\setlength{\textheight}{22cm}
\setlength{\textwidth}{174mm}
\setlength{\topmargin}{-1.5cm}

\newcommand{\beqar}[1]{\begin{eqnarray}\label{#1}}
\newcommand{\eeqar}{\end{eqnarray}}
\newcommand{\m}{\marginpar{*}}
\newcommand{\lash}[1]{\not\! #1 \,}
\newcommand{\nn}{\nonumber}
\newcommand{\D}{\partial}
\newcommand{\h}{\frac{1}{2}}
\newcommand{\g}{{\rm g}}
\newcommand{\el}{{\cal L}}
\newcommand{\A}{{\cal A}}
\newcommand{\Ka}{{\cal K}}
\newcommand{\al}{\alpha}
\newcommand{\be}{\beta}
\newcommand{\ep}{\varepsilon}
\newcommand{\ga}{\gamma}
\newcommand{\de}{\delta}
\newcommand{\De}{\Delta}
\newcommand{\et}{\eta}
\newcommand{\ka}{\vec{\kappa}}
\newcommand{\la}{\lambda}
\newcommand{\ph}{\varphi}
\newcommand{\si}{\sigma}
\newcommand{\ro}{\varrho}
\newcommand{\Ga}{\Gamma} 
\newcommand{\om}{\omega}
\newcommand{\La}{\Lambda}  
\newcommand{\tG}{\tilde{G}}
\renewcommand{\theequation}{\thesection.\arabic{equation}}

%
\def\ap#1#2#3{     {\it Ann. Phys. (NY) }{\bf #1} (19#2) #3}
\def\arnps#1#2#3{  {\it Ann. Rev. Nucl. Part. Sci. }{\bf #1} (19#2) #3}
\def\npb#1#2#3{    {\it Nucl. Phys. }{\bf B#1} (19#2) #3}
\def\plb#1#2#3{    {\it Phys. Lett. }{\bf B#1} (19#2) #3}
\def\prd#1#2#3{    {\it Phys. Rev. }{\bf D#1} (19#2) #3}
\def\prep#1#2#3{   {\it Phys. Rep. }{\bf #1} (19#2) #3}
\def\prl#1#2#3{    {\it Phys. Rev. Lett. }{\bf #1} (19#2) #3}
\def\ptp#1#2#3{    {\it Prog. Theor. Phys. }{\bf #1} (19#2) #3}
\def\rmp#1#2#3{    {\it Rev. Mod. Phys. }{\bf #1} (19#2) #3}
\def\zpc#1#2#3{    {\it Z. Phys. }{\bf C#1} (19#2) #3}
\def\mpla#1#2#3{   {\it Mod. Phys. Lett. }{\bf A#1} (19#2) #3}
\def\nc#1#2#3{     {\it Nuovo Cim. }{\bf #1} (19#2) #3}
\def\yf#1#2#3{     {\it Yad. Fiz. }{\bf #1} (19#2) #3}
\def\sjnp#1#2#3{   {\it Sov. J. Nucl. Phys. }{\bf #1} (19#2) #3}
\def\jetp#1#2#3{   {\it Sov. Phys. }{JETP }{\bf #1} (19#2) #3}
\def\jetpl#1#2#3{  {\it JETP Lett. }{\bf #1} (19#2) #3}
\def\ppsjnp#1#2#3{ {\it (Sov. J. Nucl. Phys. }{\bf #1} (19#2) #3}
\def\ppjetp#1#2#3{ {\it (Sov. Phys. JETP }{\bf #1} (19#2) #3}
\def\ppjetpl#1#2#3{{\it (JETP Lett. }{\bf #1} (19#2) #3} 
\def\zetf#1#2#3{   {\it Zh. ETF }{\bf #1}(19#2) #3}
\def\cmp#1#2#3{    {\it Comm. Math. Phys. }{\bf #1} (19#2) #3}
\def\cpc#1#2#3{    {\it Comp. Phys. Commun. }{\bf #1} (19#2) #3}
\def\dis#1#2{      {\it Dissertation, }{\sf #1 } 19#2}
\def\dip#1#2#3{    {\it Diplomarbeit, }{\sf #1 #2} 19#3 }
\def\ib#1#2#3{     {\it ibid. }{\bf #1} (19#2) #3}
\def\jpg#1#2#3{        {\it J. Phys}. {\bf G#1}#2#3}  
%

%
\def\thefootnote{\fnsymbol{footnote}} 
%
%
%
\noindent
\begin{flushright}
\parbox[t]{10em}{
TAUP-2732-2003 \\ 
{\tt hep-ph/0305150} \\
 \today }\\
\end{flushright}
\vspace{1cm}
\begin{center}
{{\LARGE  \bf   QCD Saturation in the Semi-classical Approach}\\

\vskip1cm
{\large \bf S. Bondarenko ${}^{\ast}$  
\footnotetext{ ${}^{\ast}$ \,\,Email:serg@post.tau.ac.il.},
 ~ M. ~Kozlov ${}^{\dagger}$\footnotetext{ ${}^{\dagger}$ \,\,Email:
kozlov@post.tau.ac.il
.} and E. ~Levin ${}^{\ddagger}$ \footnotetext{${}^{\ddagger}$ \,\,Email:
leving@post.tau.ac.il, levin@mail.desy.de.}}}
\vskip1cm

{\it  HEP Department}\\
{\it School of Physics and Astronomy}\\
{\it Raymond and Beverly Sackler Faculty of Exact Science}\\
{\it Tel Aviv University, Tel Aviv, 69978, Israel}\\
\vskip0.3cm

\end{center}  
\bigskip
\begin{abstract} 	
In this paper the semi-classical approach to the solution of non-linear 
evolution equation is developed. We found the solution in the 
entire kinematic region to the 
non-linear evolution equation that governs the dynamics in the high 
parton density QCD. 

The large impact parameter ($b_t$)  
behavior of 
the solution is discussed as well as  the way how to include the 
non-perturbative 
QCD corrections in this region of $b_t$. The geometrical 
scaling behavior and other properties of the solution in  
the saturation (Color Glass Condensate) 
kinematic domain   are  analyzed. We obtain the asymptotic behavior 
for the physical observables and found the unitarity bounds for 
them.

\end{abstract}

\newpage 

\def\thefootnote{\arabic{footnote}} 
\section{Introduction}
\label{sec:Introduction}
High density QCD \cite{GLR,MUQI,MV} which is dealing with the parton 
systems where the gluon 
occupation numbers are large, has entered  a new phase of it's 
development: a direct comparison with the experimental data.  A 
considerable success \cite{GW,KS,GLM,LUB,ESK,KML,KLN} has been reached in 
description of new precise 
data 
on deep inelastic scattering \cite{HERADATA} as well as in understanding 
of general features of hadron production in ion-ion collision 
\cite{RHICDATA}. The intensive theoretical  \cite{SCA,SAT,ELTHEORY}  work 
lead to effective theory in the high parton density region with 
non-linear evolution equation for dipole-dipole amplitude that 
governs 
the dynamics in this region \cite{KV}. The practical application of 
the high density QCD approach to  a comparison with the experimental 
data is based on analytic \cite{KOV,LT,IIM} and numerical 
\cite{GMS,LUB,LT,AB,LLU}  solution to the non-linear equation. 
In spite of  understanding of the main qualitative and, partly, 
quantative properties of the non-linear dynamics,
this equation has not been solved 
at arbitrary values of the impact parameters ($b_t$) 
and a number of {``ad hoc"} ansatzs were used for $b_t$-dependence.

It turns out that 
$b_t$-dependence  is a challenging problem since the non-perturbative 
corrections are important in the large $b_t$ kinematic region. There 
are two different ideas on the market how to include the 
non-perturbative large $b_t$ behavior into high density QCD 
dynamics. The first one \cite{LRREV,FIIM,BKL} claims that the 
non-perturbative corrections could be taken into account only in 
initial condition for the non-linear evolution equation, while the 
other idea demands the change of the kernel in the equation 
\cite{KW}.

The objective of this paper is to find the semi-classical solution to 
the non-linear evolution equation with a special attention to the 
large 
$b_t$ behavior of the solution.   The semi-classical approach has 
several 
advantages\cite{GLR,SCA,SAT}. Firstly, it gives simple, transparent and 
analytic 
solution to 
the non-linear equation. Secondly, this solution has a good theoretical 
accuracy in the most interesting kinematic region: at low $x$ and small 
sizes of interacting dipoles. Thirdly, this approach leads to a natural 
definition of the new saturation scale which is  the principle 
dimensional parameter that governs dynamics at high density QCD region 
\cite{GLR,MUQI,MV}. It appears in the semi-classical approach as a 
critical 
line which divides the whole set of trajectories in two parts: inside and 
outside of the saturation region. In other words, this critical line is a 
separation line between two phase: color gluon condensate (CGC) and  
the gluon liquid and the saturation scale is the order parameter for this 
phase transition.

The shortcoming of the semi-classical approach is the fact that the 
experimental data are not in the kinematic region where this approach 
has a good theoretical accuracy. However, we believe, that we can develop 
the reasonable method for the numerical solution of the non-linear 
equation only after studying  the property of the semi-classical 
approach.

In the next two sections we discuss the semi-classical approach for the 
dipole 
sizes smaller than the saturation scale ($r_t\,<\,1/Q_s(x)$ where $Q_s 
(x)$ 
is the saturation momentum).  We review the semi-classical solution 
which 
has been investigated in this region (see Refs. \cite{GLR,SCA,SAT}) 
concentrating mostly on large  $b_t$-behavior of this solution.

Section 4 is devoted to the semi-classical   approach inside of the 
saturation 
region ($r_t\, >\,1/Q_s(x)$). We suggest a semi-classical solution 
which 
describes the behavior of the dipole-dipole amplitude deeply inside of 
this domain and discuss the matching of this solution with the 
semi-classical solution at small dipole sizes.  It turns out that this 
matching occurs on the special trajectory of the non-linear equation which 
defines the saturation scale for dipole-dipole rescattering.

In section 5 we discuss the unitarity bound as well as energy behavior of 
the gluon structure function and the $\gamma^* - \gamma^*$ total cross 
section.

Section 6 gives the estimates of the accuracy of our approach. We 
calculate the first enhanced diagram as well as corrections to the 
non-linear equation that we neglected in our solution.

In the last section we summarize our results and discuss the possible 
practical applications.

\section{ General Approach}

\subsection{ The non-linear evolution equation}

The non-linear equation that governs the dynamics in the high parton 
density QCD domain can be written in the following form given by Balitsky 
and Kovchegov \cite{KV}:
\beq \label{EQ}
\frac{\partial N({\mathbf{x_{01}}},y;b_t)}{\partial\,y}\,=\, 
-\frac{2
\,C_F\,\as}{\pi} \,\ln\left( 
\frac{{\mathbf{x^2_{01}}}}{\rho^2}\right)\,\,N({\mathbf{x_{01}}},y;b_t)\
+  \frac{2
\,C_F\,\as}{\pi}\,\times 
\eeq
$$ 
\left( \int_{\rho} \, d^2 {\mathbf{x_{2}}} 
\frac{{\mathbf{x^2_{01}}}}{{\mathbf{x^2_{02}}}\,
{\mathbf{x^2_{12}}}} 
\left(\,2\,N({\mathbf{x_{02}}},y;{ \mathbf{ b_t -
\frac{1}{2}
x_{12}}})-N({\mathbf{x_{02}}},y;{ \mathbf{ b_t -
\frac{1}{2}
x_{12}}})N({\mathbf{x_{12}}},y;{ \mathbf{ b_t - \frac{1}{2}
x_{02}}})\right)\,\right) 
$$ 

\begin{figure}
\begin{center}
\epsfig{file=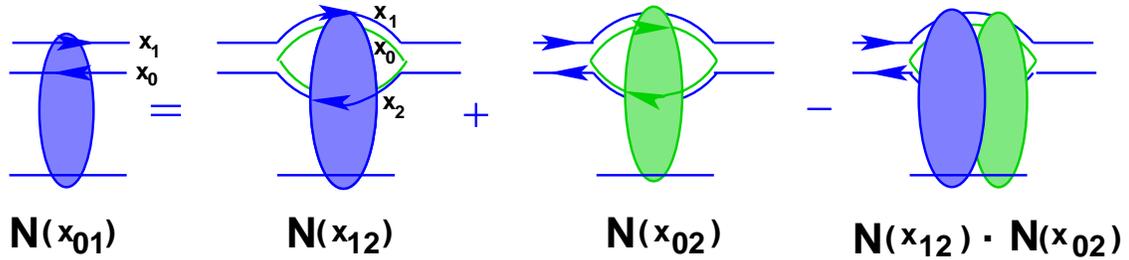,width=150mm}
\end{center}
\caption{The non-linear evolution equation.}
\label{nleq}
\end{figure}

\begin{figure}[htbp]
\begin{center}
\epsfig{file= 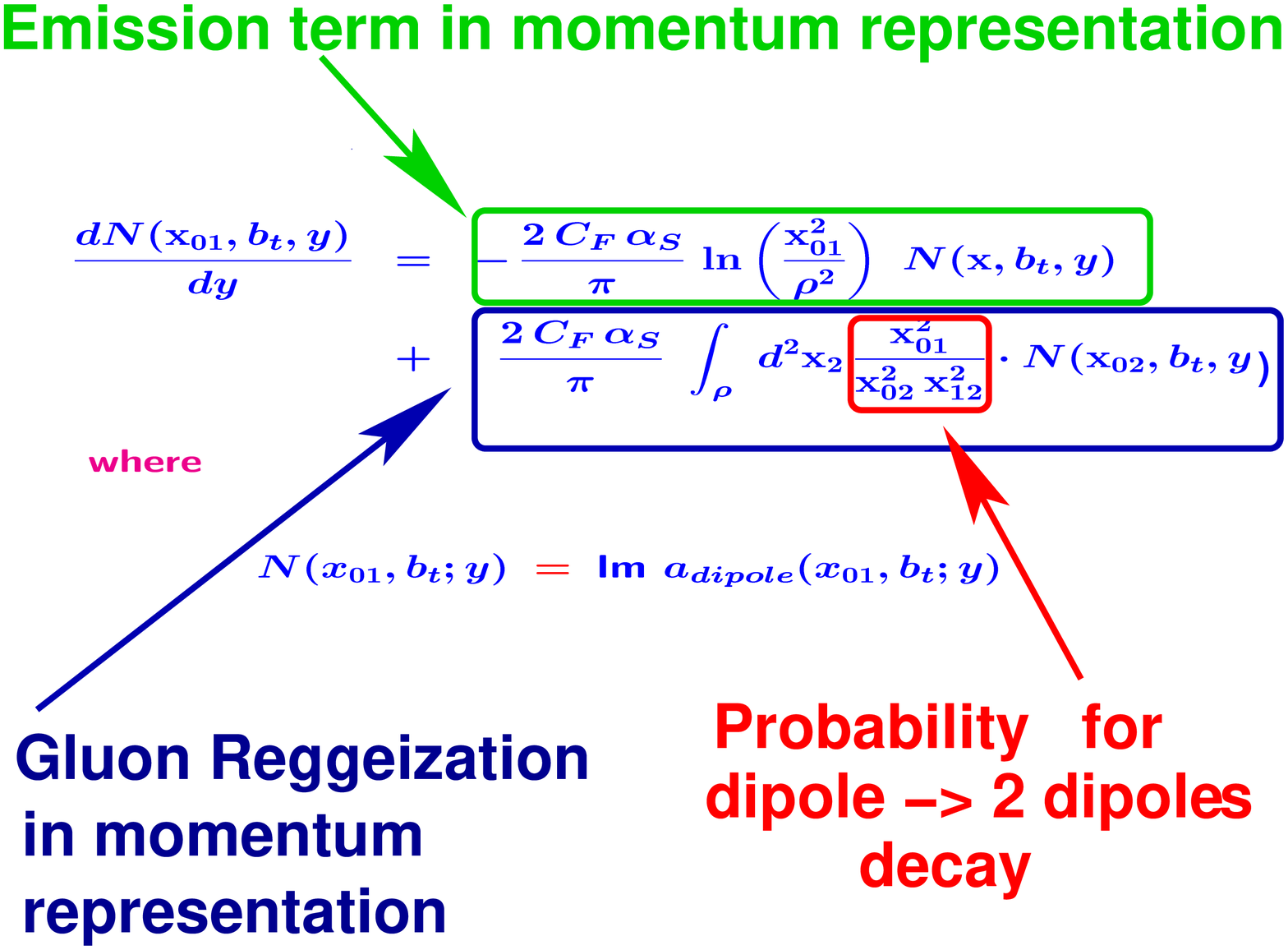,width=130mm}
\end{center}
\caption{The linear (BFKL) equation.}
\label{lneq}
\end{figure}

The meaning of \eq{EQ} is very simple and can be seen in \fig{nleq} and 
\fig{lneq}. It describes the process of dipole 
interaction as two stages process. The first stage is decay of the initial 
dipole with size $x_{01}$ into two dipole with sizes $ x_{12}$ and $
x_{02}$ with probability
 $$ |\Psi(x_{01} \,\rightarrow\,x_{02} 
\,+\,x_{12} ) |^2 \,=\,\frac{x^2_{01}}{x^2_{02}\,x^2_{12}}.$$
In the second stage two produced dipoles interact with the target. The 
non-linear part of \eq{EQ}  takes into account the Glauber corrections for 
such an interaction ( see \fig{nleq} ). The first term in \eq{EQ} stands 
for possibility for the initial dipole to interact with the target without 
decaying into two dipoles.

The equation is written in the coordinate representation which has a 
certain advantage since $N(x_{01},y;b_t)\,=\,Im\,a(x_{01},y;b_t)$ where 
$a$ is the dipole amplitude. Directly from the unitarity constraint 
follows that $N \,\leq\,1$ giving the natural asymptotic behavior  for 
$N$, namely 
$N$ tends to unity at high energies. On the other hand the non-linear term 
in \eq{EQ} contains the integration over sizes of produced dipoles. It 
means that we cannot conclude that the non-linear term is small even when 
the sizes of the initial dipoles are  very small. 

It turns out to be useful to consider \eq{EQ} in a mixed representation, 
fixing $b_t$ but introducing transverse momenta as conjugated variables 
to dipole sizes. The relation between these two representations are given 
by the following equations (see also \fig{lneq})
\footnote{The extra factor $x^2$ in \eq{EQ1} makes $\tilde{N}$
dimensionless.}:
\begin{eqnarray}
N(x, y; b_t) &=& x^2\,\int^{\infty}_0\,\,k\,dk 
\,J_0(kx)\,\tilde{N}(k,y;b_t)\,\,; \label{EQ1}\\
\tilde{N}(k,y;b_t) &=& \int^{\infty}_0\,\frac{d x}{x}\,J_0(kx)\,N(x, y; 
b_t) \,\,. \label{EQ11} 
\end{eqnarray}

In momentum representation the nonlinear term in \eq{EQ}, being the 
convolution  in the coordinate representation,  transforms to the 
product of two amplitudes,
$(N)$, at the same value of the initial transverse momentum ($k$). 
Therefore, the large value of the initial 
transverse momentum guarantees the smallness of the non-linear 
corrections. One can actually see that the non-linear term in \eq{EQ} is a 
convolution only at large value of the impact parameter $b_t$. Indeed, let 
us rewrite the non-linear term in \eq{EQ} in the momentum representation 
going also to momentum transfer instead of $b_t$.

In this case 
\beq \label{EQ3}
\tilde{N}(k,y;Q)\,=\,\int \,\frac{d^2\,b_t}{(2\,\pi)^2}\,e^{ - i 
\vec{Q}\cdot\vec{b}_t}\,\,\tilde{N}(k,y;b_t)\,\,.
\eeq

The non-linear term reduces to the form
$$
\int_{\rho} \, d^2 {\mathbf{x_{2}}}
\frac{{\mathbf{x^2_{01}}}}{{\mathbf{x^2_{02}}}\,
{\mathbf{x^2_{12}}}}N({\mathbf{x_{02}}},y;{ \mathbf{ b_t -
\frac{1}{2}
x_{12}}})N({\mathbf{x_{12}}},y;{ \mathbf{ b_t - \frac{1}{2}
x_{02}}})\,\,\longrightarrow
$$
\beq \label{NLTERM} 
\int d^2 Q\, d^2 Q' \,\,\delta(\vec{k} - \vec{k}' +\h \vec{Q} - \h \vec{Q}')
\,\tilde{N}(k,y;Q)\,\tilde{N}(k',y;Q')\,e^{\vec{Q} \cdot \vec{b}_t}\,e^{\vec{Q}' \cdot \vec{b}_t }\,\,.
\eeq
At large values of $b_t \,\gg\,1/k$ and $1/k'$ both $Q$ and $Q'$ 
are small and of the order of $1/b_t$. Neglecting terms $\h\vec{Q}$ and $\h \vec{Q}'$ in 
$\delta$-function in \eq{NLTERM}   we can reduce 
this equation  to the product of $\tilde{N}(k,y;b_t)\cdot\tilde{N}(k,y;b_t)$.

\begin{figure}[htbp]
\begin{minipage}{10.0cm}
\epsfig{file= 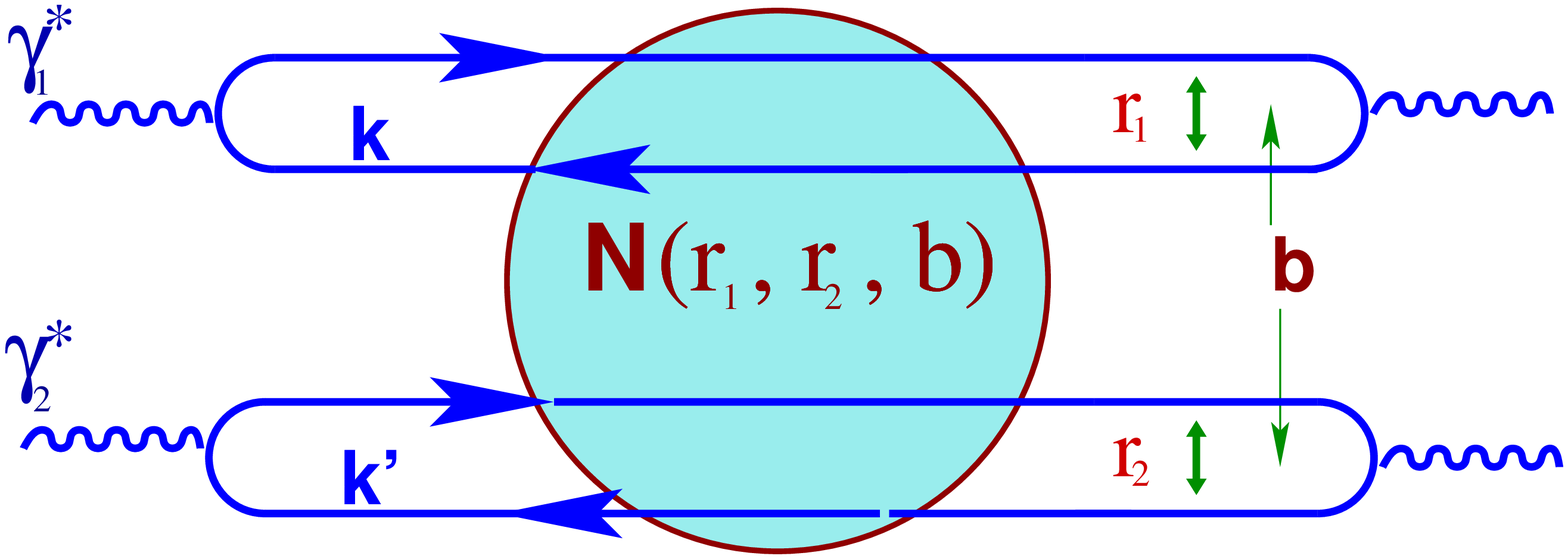,width=90mm, height=40mm}
\end{minipage}
\begin{minipage}{6.0 cm}
\caption{The picture of interaction of two  photons with virtualities
 larger than a  ``soft" scale. $\vec{k}$ and $\vec{k}'$ are 
transverse
momenta of quarks.}
\label{gaga}
\end{minipage}
\end{figure}

\begin{figure}[htbp]
\begin{tabular}{c c c}
\epsfig{file= 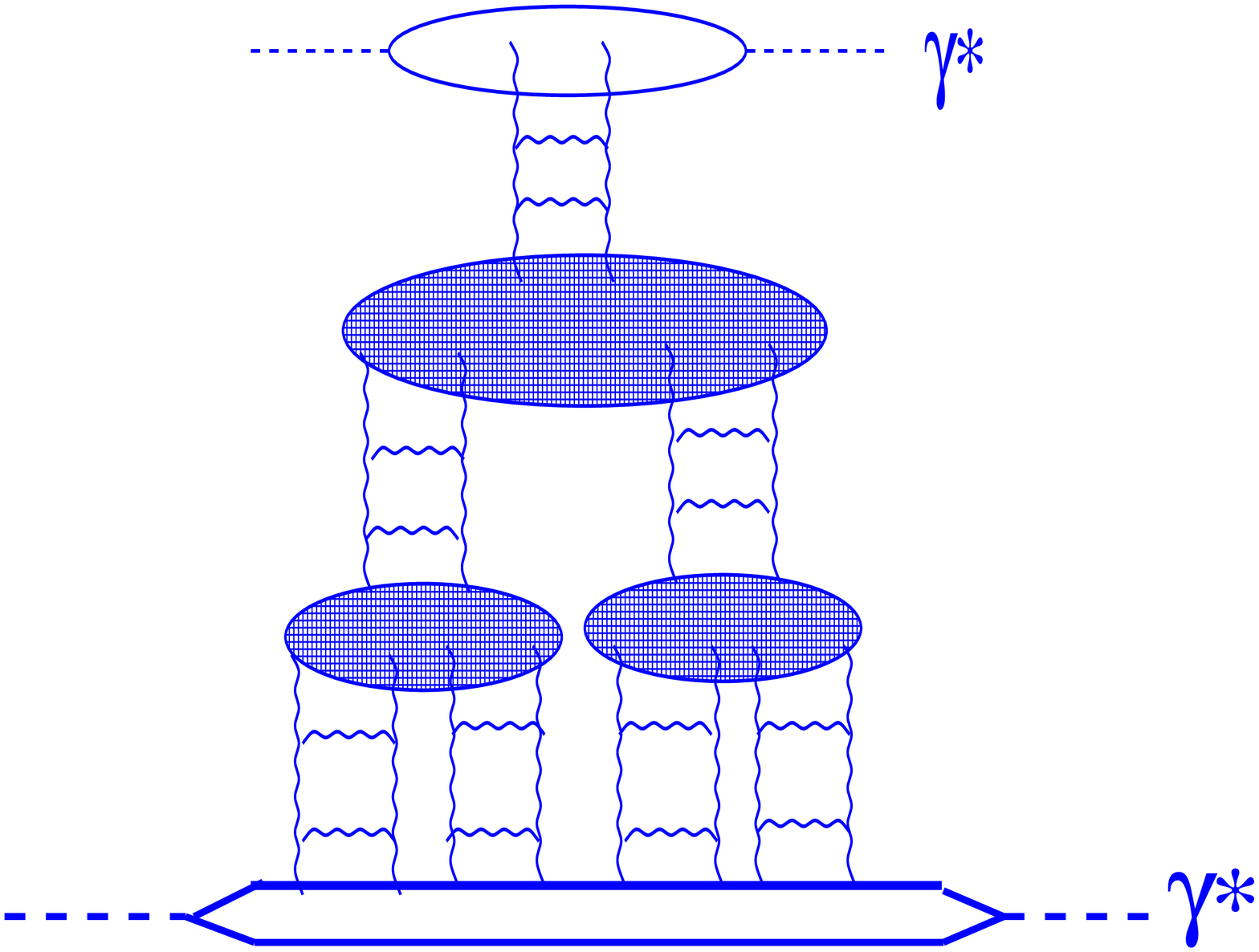,width=55mm, height=38mm} &
\epsfig{file= 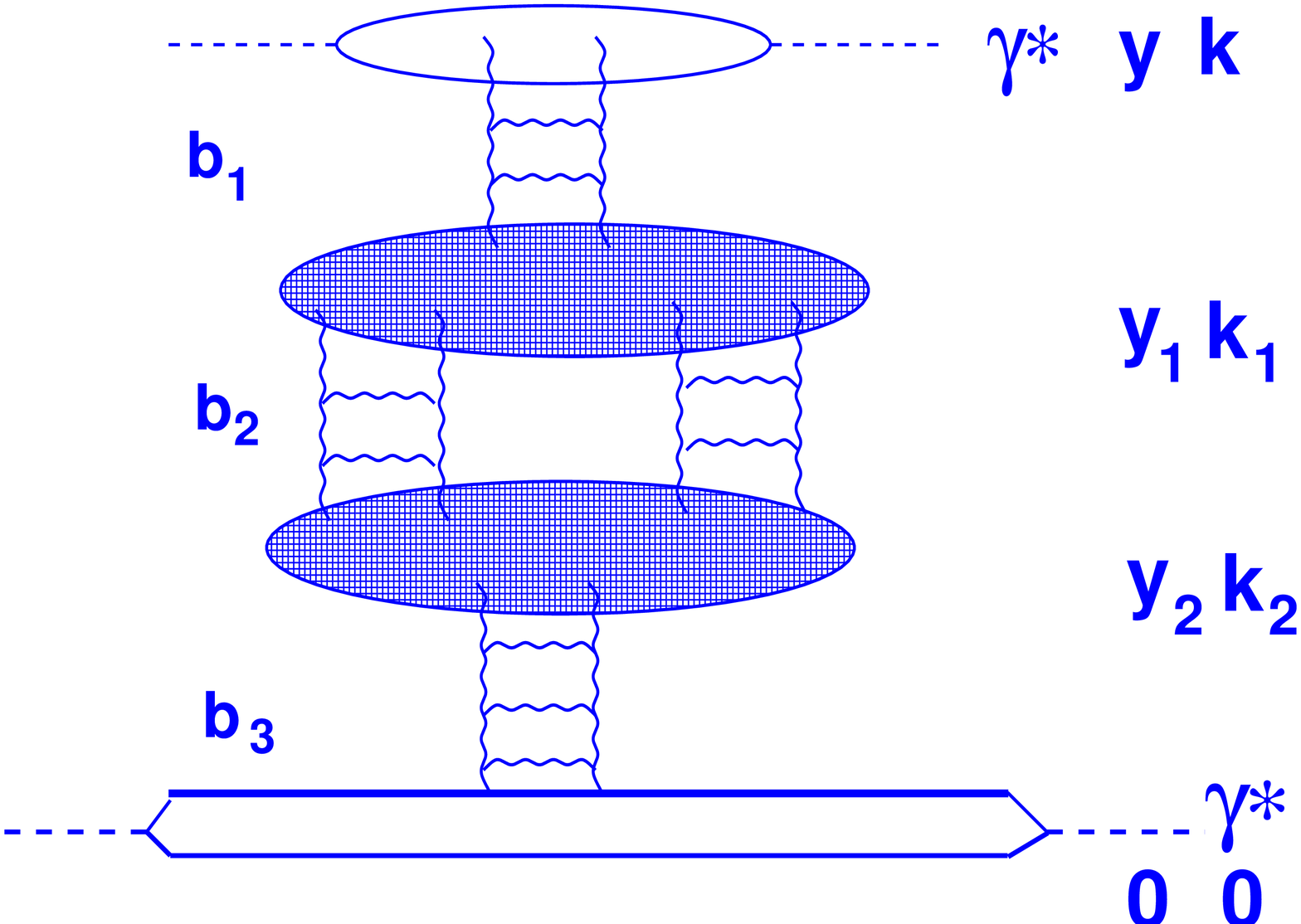,width=55mm, height=42mm}&
\epsfig{file= 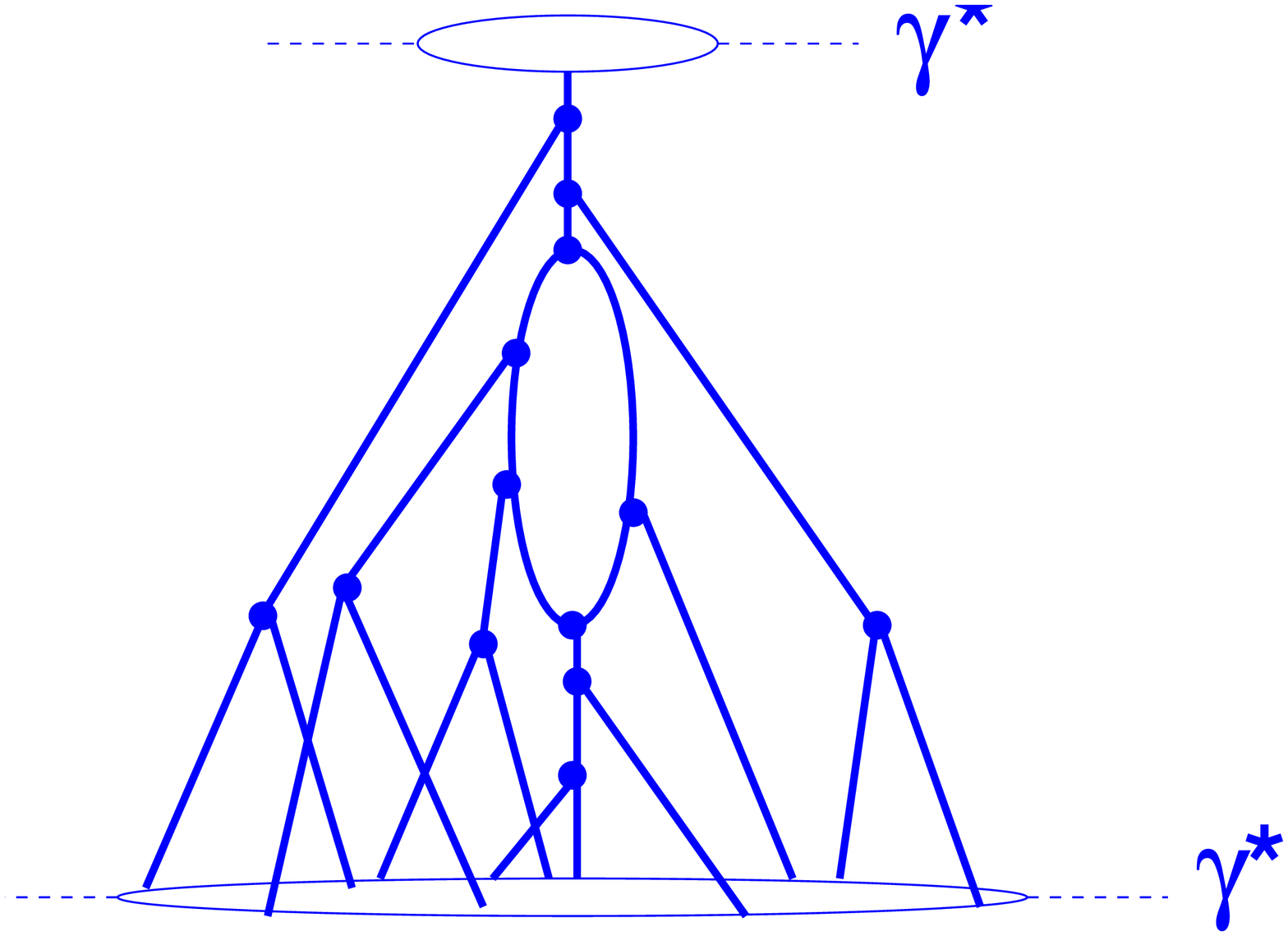,width=55mm, height=38mm}\\
\fig{fan}-a & \fig{fan}-b & \fig{fan}-c\\
\end{tabular}
\caption{``Fan" diagrams (\fig{fan} -a ), the first enhanced diagram 
(\fig{fan} 
-b )  
and the first correction to ``fan" diagrams (\fig{fan} -c )  
 for $\gamma^* - \gamma^*$ interaction. The thick 
lines in \fig{fan} -c denote the BFKL  Pomeron  .}
\label{fan}
\end{figure}

However, one can see that there is a region of integration over $Q$ , 
namely, $\vec{Q} + \vec{Q}' \,\approx\,1/b_t \,\,\ll\,\,k $ or/and 
$k'$, while $\vec{Q} - \vec{Q}'\,\,>\,\,\,k $ or/and $k'$. In the 
coordinate representation (see \eq{EQ}) this region of integration 
corresponds to $x_{12} 
\,\approx\,x_{02}\,\,\approx\,\,2\,b_t\,\,\gg\,x_{01}$. In this 
region the non-linear term can be reduced to
\beq \label{NLTERM1}
\int_{\rho} \, d^2 {\mathbf{x_{2}}}
\frac{x^2_{01}}{b^4_t} N^2(2b_t,y;\Delta  b_t)
\eeq
with $\Delta\vec{b}_t\,\,=\,\vec{b}_t\,\,-\,\,\h \vec{x}_{02}$.
In the region of large $b_t\,\,\gg\,r_2\,>\,x_{01}=r_1$ (see 
\fig{gaga} for all notations) the dipole amplitude in \eq{NLTERM1} 
describes the interaction of the dipole with the size which is much 
larger than the size of the lower dipole in \fig{gaga}. Such 
interaction cannot be described by \eq{EQ}. Indeed, the non-linear 
equation is proven only in the case when $r_1 = x_{01}\,\,\ll\,r_2$ 
since only in such conditions we can restrict ourselves by 
consideration of the ``fan" diagrams  (see \fig{fan}-a) and
only in the case of the running QCD coupling \cite{GLR}. 
\eq{EQ} is also valid  for  deep inelastic processes with 
nuclei \cite{KV} but we will not consider this process here since in 
this case the dipole amplitude does not depend on $b_t$ for $b_t \leq 
R_A$ where $R_A$ is the nucleus radius \footnote{These two cases 
when we can trust the non-linear equation (see \eq{EQ}) were 
discussed in the review talk of M. Ryskin at DIS'03 conference. We 
refer to this talk for more details.}.  Therefore, we should 
consider this region of integration as a correction to the non-linear 
equation together with the enhanced  diagrams of \fig{fan}-b.

Our strategy will be the following: we will solve the non-linear 
equation assuming that $Q'$ and $Q$ $ \approx 1/b_t \,\ll\,k $ and 
$k'$ and, using this solution, we will come back to \eq{EQ} and 
we will discuss the contribution of the region where  $\vec{Q} + 
\vec{Q}' \,\approx\,1/b_t \,\,\ll\,\,k $ or/and
$k'$., while $Q' \approx Q\approx k$ or $k'$.
  
Finally, \eq{EQ} can be written in the form (see Ref. \cite{LT} for 
details )
\beq \label{EQK}
\frac{\partial \tilde{N}(k,y;b_t)}{ \partial 
\,y}\,\,=\,\,\bas\,\left(\,\hat{\chi}(\hat{\gamma}(k))\tilde{N}(k,y;b_t)\,\,-
\tilde{N}^2(k,y;b_t)\,\right)\,\,;
\eeq
where $\hat{\gamma}(k)$ is an operator  corresponding to the anomalous 
dimension of the gluon structure function and it is equal to
\beq \label{OG}
\hat{\gamma}(k)\,\,=\,\,1\,\,+\,\,\frac{\partial}{\partial\,\xi}
\eeq

The form of the first term on the r.h.s. of \eq{EQK} as well as 
definition of the variable $\xi$,  we will discuss 
in the next section. Here, we would like only to draw your attention 
to the fact that the emission of the gluons in the BFKL equation 
is described by the first tern in the r.h.s. of \eq{EQ} (see 
\fig{lneq}) which enters at the same value of the impact parameter as 
the l.h.s. of the equation.

Function $\chi(\gamma)$ is an eigen value of the BFKL   equation 
\cite{BFKL}
\beq \label{BFKLK}
\chi (\gamma)\,\,=\,\,2\,\psi(1)\,-\,\psi(1 - 
\gamma)\,-\,\psi(\gamma)\,\,.
\eeq

\eq{EQK} we will solve in the semi-classical approach.
 
\subsection{ The solution to the BFKL equation at large 
$\mathbf{b}_{\mathbf{t}}$}

This solution has been discussed (see Refs. 
\cite{BFLLRW,NP,LI,LIP,BKL} ) and at large $b_t$ it has a  form:
\beq \label{BFKLLB}
\tilde{N}(k,k',y;b_t)\,\,=\,\,\int\,\,\frac{d 
\gamma}{2\,\pi\,i}\phi_{in}(k',\gamma; 
b_t)\,e^{\bas\,\chi(\gamma)\,y\,\,- \,\,( 1\,-\,\gamma)\,\xi}
\eeq
where $\xi \,=\,\ln(b^4_t\,k^2\,k'^2)$ and $\phi_{in}$ is determined by 
the 
initial conditions.

\begin{figure}[htbp]
\begin{minipage}{12.0cm}
\epsfig{file= 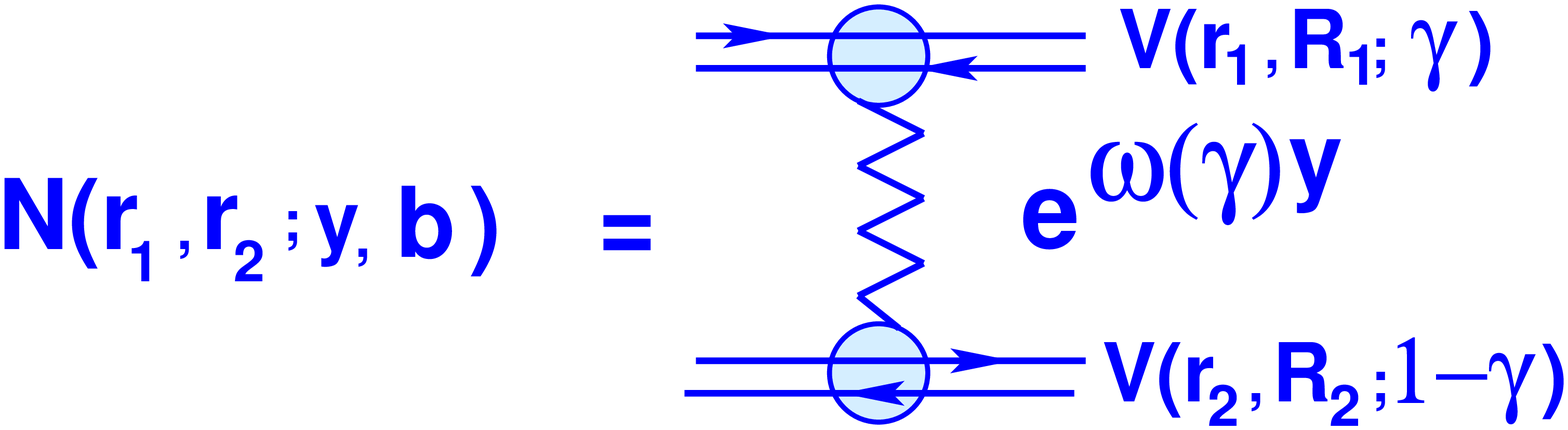,width=110mm}
\end{minipage}
\begin{minipage}{4.5 cm}
\caption{The BFKL Pomeron contribution at fixed $b_t$.}
\label{bfkl}
\end{minipage}
\end{figure}

It is easy to see that \eq{BFKLLB} is a solution to the \eq{EQ} 
without the non-linear term. Since the form of solution given by 
\eq{BFKLLB} is very important both for a derivation of the linear 
part of \eq{EQ} and for  understanding the value of the typical 
dipole sizes in the non-linear part of the equation we will discuss 
this solution in some details follow Refs. \cite{NP,LI,LIP}. The 
general solution to the BFKL equation in the coordinate 
representation was derived in Ref.\cite{LIP}, namely (see 
\fig{bfkl}),
\beq \label{BFKL1}
N( r_1,r_2;y, b_t)\,\,=
\eeq
$$
\int \frac{d \gamma}{2\,\pi\,i}\,\phi_{in}(\gamma;r_1)
\,\,d^2\, R_1 \,\,d^2\,R_2\,\delta(\vec{R}_1 - \vec{R}_2 - 
\vec{b}_t)\,
e^{\omega(\gamma)\,y}
\,V(r_1,R_1;\gamma)\,V(r_2,R_2;1 -\gamma)
$$

with
\beq \label{OMEGA}
\omega(\gamma)\,\,=\,\,\bas \chi(\gamma)\,\,  
\eeq
and
\beq \label{V}
V(r_i,R_i;\gamma)\,\,=\,\,\left( \,\frac{r^2_{i}}{(\vec{R}_i
\,+\,\frac{1}{2}\vec{r}_{i})^2\,\,
(\vec{R}_i\,-\,\frac{1}{2}\vec{r}_{i})^2}\,\right)^{1 - \gamma}\,\,.
\eeq

The integration over $R_1$ in \eq{BFKL1} was performed in  Refs. 
\cite{NP,LI} with the result:
$$
\int \,\,d^2\,R_1\,V(r_{1},R_1; \gamma)\,\, V(r_{1},|\vec{R}_1
\,-\,\vec{b}_t|; 1 - \gamma)\,=
$$
$$
\,\frac{ (\gamma - \h)^2}{( 
\gamma (1 - \gamma)
)^2}(\,c_1\,x^\gamma\,{x^*}^\gamma\,F(\gamma,\gamma,2\gamma,x)\,
F(\gamma,\gamma,2\gamma,x^*) 
\,+
$$
\beq \label{A1} 
+\,\, c_2 x^{1 -
\gamma}{x^*}^{1-\gamma}
F(1 - \gamma,1 -\gamma,2 - 2\gamma,x)\,F(1 - \gamma,1 -\gamma,2 
-2\gamma,x^*) \,)\,,
\eeq
where $F$ is hypergeometric function \cite{RY} and
$x \,x^*$ is equal to \cite{NP,LI}
\beq \label{A5}
x\,x^*\,\,=\,\,\frac{r^2_{1,t}\,r^2_{2,t}}{(\vec{b} - 
z_1\,\vec{r}_{1,t}
-\bar{z}_2\,\vec{r}_{2,t})^2
\,(\vec{b} - \bar{z}_1\,\vec{r}_{1,t} - z_2\,\vec{r}_{2,t})^2}
\eeq
and $c_1$ and $c_2$ are function of $\gamma$.

At large values of $b_t \gg \,r_1 $  and $r_2$ the argument $x\,x^*$ is small, 
namely
\beq \label{XX}
x\,x^*\,\,=\,\,\frac{r^2_1\,r^2_2}{b^4_t}
\eeq

It is interesting to notice that $x\,x^*$ turns out to be small and 
simple also for $\,b_t\,<\,r_2$ with $\,r_1\,<\,r_2$. 
Indeed, for such values of $b_t$ 
\beq \label{XX0}
x\,x^*\,\,=\,\,\frac{r^2_1\,}{z^2_2\,\bar{z}^2_2\,r^2_2}
\eeq

Expanding \eq{A1} and rewriting the answer in the  transverse 
momentum 
representation given by \eq{EQ1} and \eq{EQ11} we reduce the general 
solution of \eq{A1}  to  \eq{BFKLLB}. 
Using \eq{BFKLLB} one can see that the linear part of \eq{EQK} indeed 
gives the BFKL equation. Since all derivative in \eq{EQK} are taken 
with respect to $\xi$ we can consider $\phi_{in}$ in \eq{EQK} as an 
arbitrary function of $b_t$.

The exact form of the initial condition depends on the 
particular reaction. We choose as an instructive example the virtual 
photon-photon scattering (see \fig{gaga}).

The advantage of this process is the fact that we know the initial 
condition which is under control of pQCD except large $b_t$ behavior.
The problem of large $b_t$ behavior of the Born amplitude is  addressed 
 in Ref. \cite{BKL} and it is shown that $\phi_{in}$ can 
be chosen in the form
\beq \label{INPHI}
\phi_{in}(k',y;b_t) \,=\,\pi \as^2 \frac{N^2_c - 
1}{3\,N^2_c}\,\,(m_{\pi}\,b_t)^4 \,K_4( 2 \,m_{\pi}\,b_t)\,\frac{1}{\gamma}
\eeq
to satisfy both the $1/b^4_t$ of Born approximation in perturbative 
QCD, namely,
\beq \label{INPHIPER}
\phi_{in}(k',y;b_t) \,=\,\pi \as^2 \frac{N^2_c -
1}{48\,N^2_c}\,\,\,\frac{1}{\gamma}
\eeq 
and the non-perturbative $e^{-\,2 \,m_\pi\,b_t}$ behavior at large $b_t 
\,\gg 
\,1/(2\,m_{\pi} \,b_t)$. 

\subsection{Semi-classical approach}

In semi-classical approach we are looking for a solution in the form:
\beq \label{SC1}
N(k,k',y;b_t)\,=\,e^{\omega(\xi,y,b_t)\,y\,\,-\,\,( 
\,1 \,-\,\gamma(\xi,y,b_t)\,)\,\xi\,\,\,
+\,\,\beta(k',b_t)}
\eeq
where $\omega(\xi,y,b_t)$ and $\gamma(\xi,y,b_t)$ are smooth functions of 
$y$ and $\xi$: $d \omega(\xi,y,b_t)/d y\,\,\ll\,\,\omega^2(\xi,y,b_t)$,
$d \omega(\xi,y,b_t)/d \xi 
\,\,\ll\,\,\omega(\xi,y,b_t)\,( 1 - \gamma(\xi,y,b_t))$, $d 
\gamma(\xi,y,b_t)/d 
\xi \,\,\ll\,\,(\,1\,-\,\gamma\,)^2(\xi,y,b_t)$ and

 $d 
\gamma(\xi,y,b_t)/d y 
\,\,\ll\,\,\omega(\xi,y,b_t)\,( 1 - \gamma(\xi,y,b_t))$
\footnote{The form of \eq{SC1} is obtained from the solution
of the linear BFKL equation, that has been discussed  above.}.

Assuming \eq{SC1} we can use the method of characteristic (see, for 
example,  
Ref. \cite{MC}) to solve the 
non-linear equation. For equation in the form
\beq \label{SC2}
F(y,\xi,S,\gamma,\omega)=0
\eeq
where $S = \omega \,y + \gamma \xi  + \beta $, we can introduce the set 
of 
characteristic lines on which  $\xi(t), y(t), S(t),$ $ \omega(t),$ $ 
\gamma(t)$ are  
functions of variable $t$ (artificial time), which satisfy the following 
equations:
\begin{eqnarray}
(1.)\,\,\,\,\,\,\,\frac{d\,\xi}{d\,t}\,\,=\,\,F_{\gamma}\,\,,
\,\,\,\,\,\,\,\,\,\,\,\,(2.)\,\,\,\,\,\,\,
\frac{d\,y}{d\,t}\,\,=\,\,F_{\omega}\,\,,& &(3.)\,\,\,\,\,\,\, 
\frac{d\,S}{d\,t}\,\,=\,\,\gamma\,F_{\gamma}\,+\,\omega\,F_{\omega}\,\,, 
\nonumber \\
(4.)\,\,\,\,\,\,\,\frac{d\,\gamma}{d\,t}\,\,=\,\,- 
(\,F_{\xi}\,+\,\gamma\,F_{S}\,)\,\,, & & (5.)\,\,\,\,\,\,\, 
\frac{d\,\omega}{d\,t}\,\,=\,\,- \,(\,F_{y}  \,+\,\omega\,F_S\,)\,\,, 
\label{SCEQ}
\end{eqnarray}
where $F_y = \frac{\partial F(y,\xi,S,\gamma,\omega)}{\partial y}$ etc.

We can reduce the master equation (see \eq{EQ} ) in semi-classical 
approach to the form of \eq{SC2}, namely,
\beq \label{NLEQSC}
\omega\,\,-\,\,\bas\,\chi(\gamma) \,\,+\,\, \bas\,e^S\,\,=\,\,0\,.
\eeq

Using \eq{SC1} and \eq{NLEQSC} we can write    
the set of equations  (see \eq{SCEQ})  in the form:
\begin{eqnarray}
(1.)\,\,\,\,\,\,\,\frac{d\,\xi}{d\,t}\,\,=\,\, -\,\bas 
\,\frac{d\,\chi(\gamma)}{d\ \gamma} 
\,\,\,,\,\,\,\,\,\,\,\,\,\,\,\,(2.)\,\,\,\,\,\,\,\,
\frac{d\,y}{d\,t}\,\,=\,\,1\,\,,& &(3.)\,\,\,\,\,\,\,\,
\frac{d\,S}{d\,t}\,\,=\,\,
\bas\,(1 - \gamma)\,\frac{d\,\chi(\gamma)}{d\,\gamma} 
\,+\,\omega\,\,\,, \nonumber \\
(4.)\,\,\,\,\,\,\,\,  
\frac{d\,\gamma}{d\,t}\,\,=\,\,\,\,\bas\,(\,1\,-\, 
\gamma\,)\,e^{S}\,\,, & &(5.)\,\,\,\,\,\,\,\,  
\frac{d\,\omega}{d\,t}\,\,=\,\,- \,\,\bas\,\omega\,e^{S}\,\,\,,
\label{SCEQ1}
\end{eqnarray}
We will solve these equations in the next section, but before doing this 
we are going to illustrate the method of characteristic solving the linear 
equation neglecting the non-linear tern in \eq{EQ}. In semi-classical 
approach the linear equation has a form:
\beq \label{LEQ1}
\omega\,\,=\,\,\bas\,\chi(\gamma)\,\,.
\eeq

 For linear equation 
the set of \eq{SCEQ1} reduces to two equations:
\beq \label{SCEQ2}
\frac{d\,\xi}{d\,y}\,\,=\,\,-\,\bas
\,\frac{d\,\chi(\gamma)}{d\,\gamma} \,\,\,,\,\,\,\,\,\,\,\,\,\,\,\, 
\frac{d\,S}{d\,y}\,\,=\,\,
\bas\,( 1 - \gamma)\,\frac{d\,\chi(\gamma)}{d\,\gamma}\,\,+\,\,\omega\,\,,
\eeq
with both $\gamma$ and $\omega$ being constant as function of $y$.

The solution is very simple, namely,
\beq \label{LEQ2}
S\,\,=\,\,\chi(\gamma_S)\,(y\,-\,y_0)\,\,-\,\,( 1 - \gamma_S)\,(\xi - 
\xi_0)\,+\,\beta(b_t)
\eeq
with $\gamma_S$ given by 
\beq \label{LEQ3}
\xi \,-\,\xi_0 \,\,=\,\,-\,\bas 
\,\frac{d\,\chi(\gamma)}{d\,\gamma}|_{\gamma = \gamma_S}\,(y\,-\,y_0)
\eeq

Comparing \eq{LEQ2} and \eq{LEQ3} with the solution to the BFKL equation 
of \eq{BFKLLB} one can see that the semi-classical approach is the 
exact 
solution of the linear (BFKL) equation in which the contour integral over 
$\gamma$ is taken by the steepest decent method. \eq{LEQ3} is the equation 
for the saddle point value of $\gamma = \gamma_S$. 
The accuracy of the semi-classical approach is  even worse than one 
for the 
steepest decent method since we cannot guarantee the pre-exponential 
factor in the semi-classical calculation which appears in the 
steepest 
decent method. However, the semi-classical approach has a great 
advantage 
since it allows us to treat the non-linear equation within the same 
framework as the linear one without major complications.

To solve \eq{SCEQ1} we need to find out the initial conditions for this 
set of equations, which we derive from the Glauber-Mueller formula for  
$\gamma^* - \gamma^*$ scattering  ( see Ref. \cite{BKL} ), namely:
\beq \label{GMBA}
N(x,y=y_0,;b_t) \,=\, \left(\, 1 - 
e^{-\,N^{BA}(x,y=y_0,b_t)\,} \right)
\eeq
where $N^{BA}$ is the dipole amplitude in the Born approximation.
It has been calculated in Ref. \cite{BKL}
\beq \label{XSBA}
N^{BA}(x,y=y_0,b_t)\,\,=\,\,\pi\,\as^2 \frac{N^2_c - 1}{3\,N^2_c}\,( 
m_{\pi}\,b_t)^4\,K_4(2\,m_\pi\,b_t)\,e^{\tilde{\xi}}\,\equiv\,\frac{1}{4}\, 
\tilde{\tau}(y=y_0;b_t)\,e^{\tilde{\xi}}\,,
\eeq
where the variable $\tilde{\xi} $ is defined as
\beq \label{XI}
\tilde{\xi}\,\,=\,\,\ln \left(\frac{r^2_{t,1}\,r^2_{2,t}}{( 
b^2_t\,\,+\,\,\frac{1}{4} r^2_{2,t})^2}\right)\,\,.
\eeq
Comparing \eq{XI} with \eq{XX} and \eq{XX0} we notice that
$\tilde{\xi}\,=\,\ln (x\,x^*)$ with $z_2 = - \bar{z}_2 = \h$. We put 
these values of $z_2$ just for the sake of simplicity.

One can see that $\tilde{\xi}$ in the coordinate representation can be 
replaced by $ - \xi + 4\ln 2$ with $\xi$
\beq \label{XIM}
\xi\,\,\,=\,\,\,\ln \left(\frac{k^2 
\,(\,k'^2\,b^2_t\,+\,1\,)^2}{k'^2}\right)\,\,.
\eeq
At large value of $b_t$ we have the same definition of $\xi$ that has been 
used but \eq{XIM} gives us 
 a generalization which allows  us to treat the case of $b_t = 0$.

\begin{figure}[htbp]
\begin{minipage}{10.0cm}
\epsfig{file= 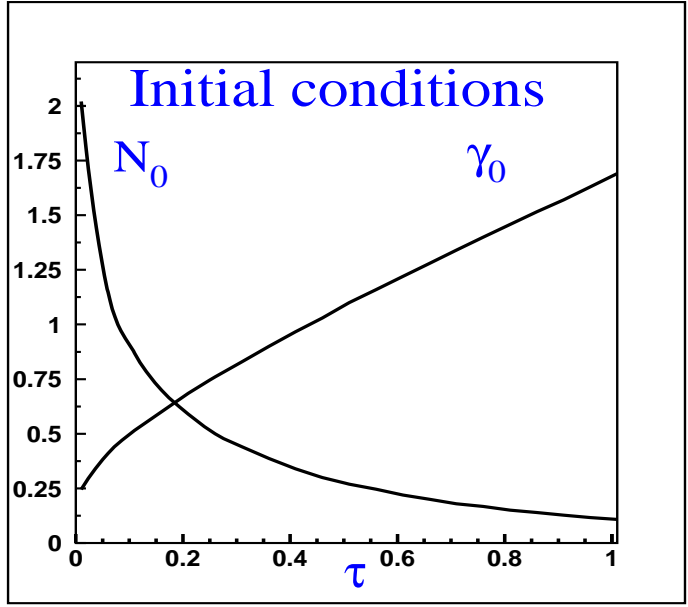,width=90mm, height=80mm}
\end{minipage}
\begin{minipage}{6.0 cm}
\caption{Initial conditions for dipole-dipole amplitude $N_0$ and 
anomalous dimension $\gamma_0$ at fixed $y=y_0$ as function of $\tau$
from Glauber-Mueller formula for $\gamma^* - \gamma^*$ scattering 
\cite{BKL}
.} 
\label{incon}
\end{minipage}
\end{figure}

Using \eq{EQ1} and \eq{EQ11} we can find the initial condition in the 
momentum representation:
\beq \label{INCONMR}
\tilde{N}_0(\xi,y=y_0;b_t)\,\,=\,\,\h\,\Gamma_0\left(\tau(\xi;b_t)\,=\,  
4\,\,e^{- 
\xi}\,\tilde{\tau}(y=y_0;b_t)\right)
\eeq
where $\Gamma_0$  is incomplete Euler gamma function \cite{RY}  of 
zeroth order. The argument $\tau(\xi;b_t)$ in \eq{INCONMR} 
can be rewritten as
\beq \label{ARG}
\tau(\xi;b_t)\,\,\,=\,\,\,\frac{Q^2_s(y=y_0,k'^2;b_t)}{k^2}\,
\tau_{0,cr}\,\,,
\eeq
with 
\beq \label{QSY0}
Q^2_s(y=y_0,k'^2;b_t)\,\,=
\,4\,\tilde{\tau}\,\tau_{0,cr}^{-1}\,k'^2\,(\,k'^2\,b^2_t\,+\,1)^{-2}\,\,
\eeq
$$
\stackrel{b_t \,>\,1/k'}{\longrightarrow} 
\,\,\,\frac{1}{k'^2\,b^4_t\,\tau_{0,cr}}\,\pi\,\alpha^{2}_{S}\,
\frac{N^{2}_{c}-1}{3\,N^{2}_{c}}\,
(2m_{\pi}\,b_t)^4\,K_4(2\,m_\pi\,b_t)\,
$$
where we define $\tau_{0,cr}$ as a value of $\tau$ at $\,y\,=\,y_0\,$
on the critical trajectory for  $\xi\,=\,\xi_{cr}(\gamma_{cr})$, see next
section. The behavior of $\tilde{N}_0$ and $\gamma_0 - 1  = 
S_\xi $ at $y = y_0$ which follow from \eq{INCONMR} is shown by \fig{incon} .
  
\begin{figure}[htbp]
\begin{minipage}{10.0cm}
\epsfig{file= 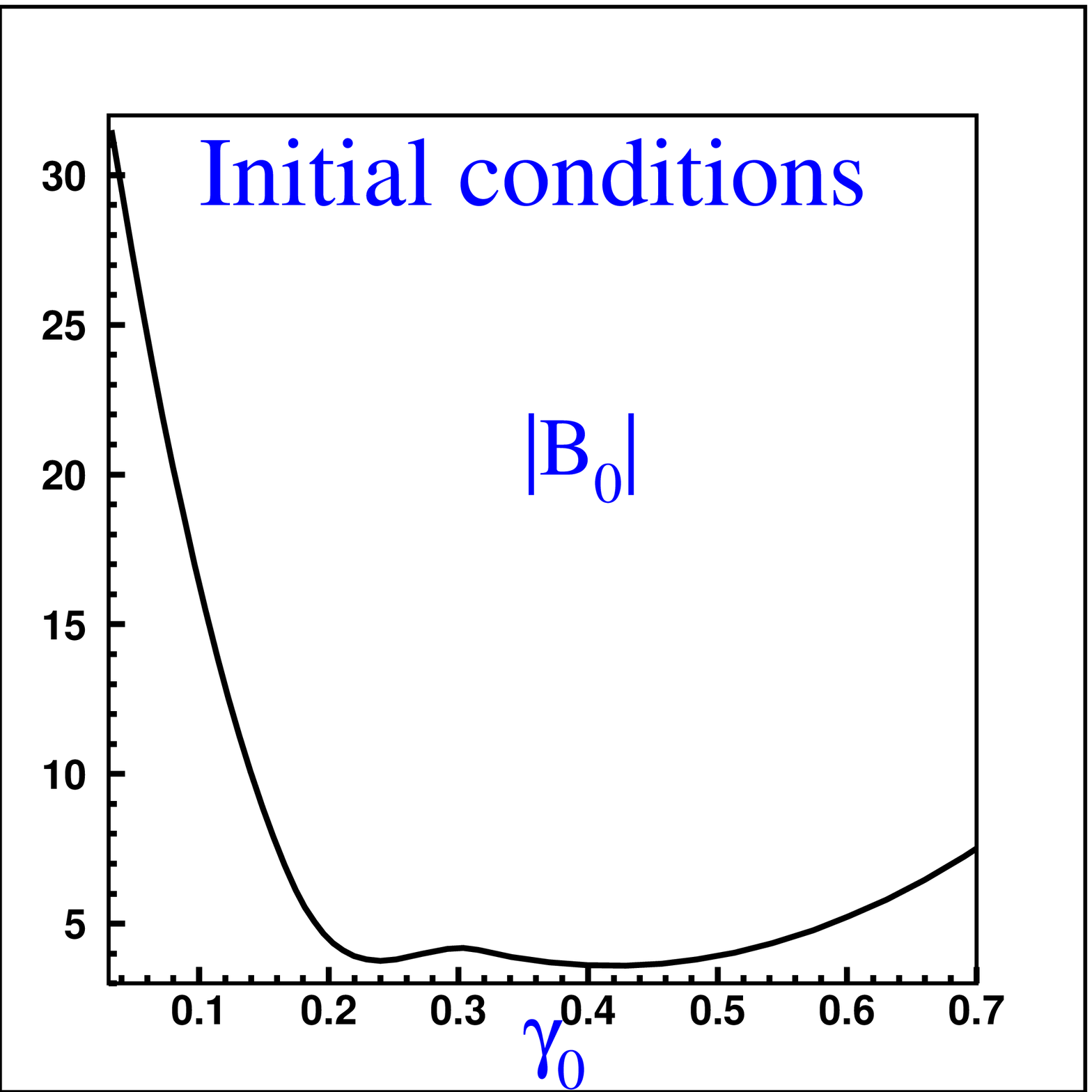,width=90mm, height=80mm}
\end{minipage}
\begin{minipage}{6.0 cm}
\caption{Initial conditions for $B_0$ given by \eq{RATB}
.}
\label{rat}
\end{minipage}
\end{figure}

As one will see below we need also to know   a ratio 
\beq \label{RATB}
B_0\,\, = \,\,\frac{\,\,\chi(\gamma_0)\,\,-\,\,\frac{N_0}{\bas}}{1 
\,-\,\gamma_0}
\eeq
for finding the solution. This ratio is shown in \fig{rat}.

\section{Solution at $\mathbf{\xi}\,\,\mathbf{ > }\,\,
\mathbf{\xi}_{\mathbf{sat}}$\,} 

\subsection{Fixed QCD coupling}

\subsubsection{Solution}

As has been discussed, see \cite{GLR,SCA,SAT}, in this kinematic region 
we can safely use the semi-classical approach for sufficiently large 
values 
of $\xi$. In this section we will find this semi-classical solution 
and 
will show, that the main qualitative feature of it, is the existence 
of 
special, critical trajectory which determines the saturation scale.

From equations 4 and 5 of \eq{SCEQ1} one can see that
\beq \label{OGEQ}
\frac{d \omega}{d \gamma}\,\,=\,\,- \,\,\frac{\omega}{1\,-\,\gamma}
\eeq
\eq{OGEQ} has a very transparent physics since it means that the phase 
and group velocity of wave package, defined by \eq{SC1}, are equal.

\eq{OGEQ} has an obvious solution 
\beq \label{OSOL}
 \omega\,=\,-\as\, B_0\,(\,1\,-\,\gamma\,)
\eeq
where $B_0$ is a constant which has to be determined from the initial 
condition (see \fig{rat} and \eq{RATB}). Using \eq{OSOL}, \eq{NLEQSC} and 
\eq{SCEQ1}- 4  
we obtain that
\beq \label{SOL1}
\frac{d\,\gamma}{d \,y}\,\, =\,\,\as\,\,(\,1\,-\,\gamma\,)\,\left(\, 
B_0\,(\,1\,-\,\gamma\,) \,\,+\,\,\,\chi(\gamma)\,\right)
\eeq

However, before solving \eq{SOL1}, we  would like to draw your attention 
to the fact that \eq{OGEQ} has itself a very interesting solution if we 
assume that non-linear corrections in \eq{NLEQSC} are small but they  
are valuable in \eq{SCEQ1}-4 and \eq{SCEQ1}-5. In other words, the 
non-linear term in the master equation is small in comparison with 
the linear one on this special trajectory but the non-linear 
contributions is essential in the equations for $S_\xi$ and $S_y$ 
dependence on $y$. 
In this case \eq{OGEQ} reduces to \beq 
\label{GAMMACR}
\frac{d\,\chi(\gamma )}{d\,\gamma}\,\,=\,\,-\,\frac{\chi(\gamma)}{1 \,- 
\,\gamma }\,\,.
\eeq
The solution to this equation is $\gamma = \gamma_{cr} \,\approx\,0.37$ 
\footnote{\eq{GAMMACR} has been derived in Ref.\cite{GLR} and 
  has been discussed in detail in Refs. \cite{BALE,MUT,TR}. 
$1\,-\,\gamma_{cr} 
$ is called  $k_0$ in Ref.\cite{GLR}, $\tilde{\gamma}_{crit}$ in Ref. 
\cite{BALE}, 
$\lambda_0$ in Refs.\cite{MUT,TR} and  $\gamma_0$ in Ref. 
\cite{BKL}.}.

The form of the trajectory is clear from \eq{SCEQ1}-1, namely
\beq \label{CRITTR}
\xi_{cr}\,\,=\,\,-\,\,\as\,\frac{d 
\,\chi(\gamma_{cr})}{d\,\gamma_{cr}}\,(y - y_0) \,\,+\,\,\xi_0(b_t).
\eeq
The value of the dipole amplitude $N$ we can find using \eq{SCEQ1}-3
which can be reduced to
\beq \label{CRITTR1}
\frac{d S}{d y} = - \as\,e^{S}\,\,,
\eeq
 which has the solution
\beq \label{CRITTR2}
N\,\,=\,\,\frac{N_0}{N_0\,\as\,(y - y_0)\,\,+\,\,1}
\eeq
Since this solution falls down at large  $y$ all our assumptions are 
self-consistent at least at $\as\,(y - y_0) \,\gg\,1$. The trajectory 
at small values of $\as\,(y - y_0)$
 should be found from \eq{SOL1}.

\subsubsection{Numerical solution}

To find the solution to the master equation we need to solve
\eq{SOL1} and obtain $\gamma(y)$ as a function of $y$, substitute this 
function into \eq{NLEQSC} and find out $N(y)$ .These two functions 
$\gamma(y)$ and $N(y)$ will determine the solution on the trajectory that 
is given by \eq{SCEQ1}-1.

\begin{figure}[htbp]
\begin{minipage}{11.0cm}
\epsfig{file= 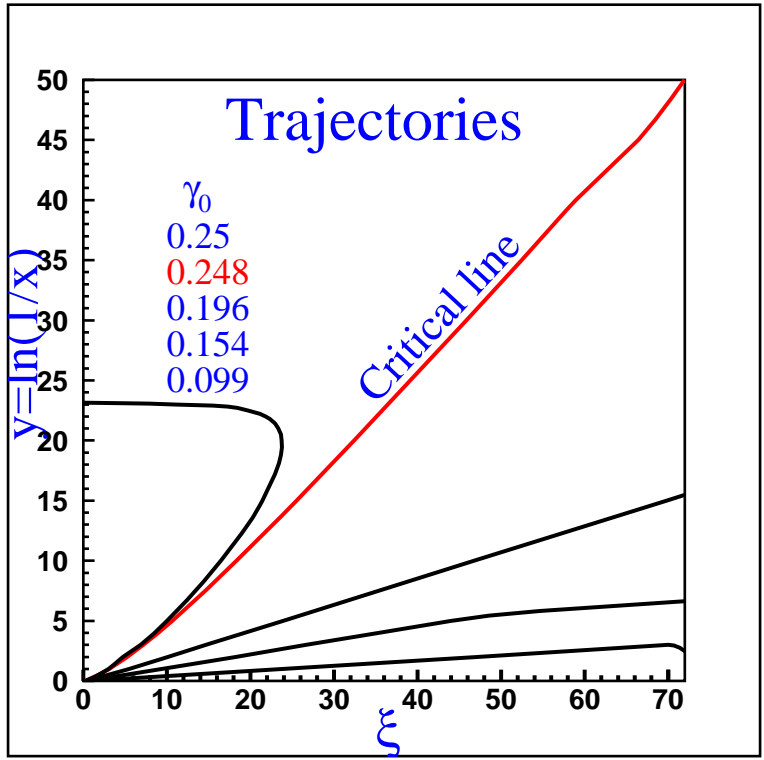,width=100mm}
\end{minipage}
\begin{minipage}{5.0 cm}
\caption{Trajectories for the non-linear evolution equation ( see \eq{EQ}) 
in semi-classical approach 
at $b_t = 0$.
.}
\label{traj}
\end{minipage}
\end{figure}

One can see from \fig{traj} that we can divide all possible trajectories 
in two part with initial $\gamma_0$ larger or smaller than $\gamma_0 
=\gamma_{0,cr}$. From $\gamma_0 =\gamma_{0,cr}$ starts the critical 
trajectory on which $\gamma(y) \,\,\rightarrow \gamma_{cr}$ at large $y$. 
The trajectories to the right of the critical one are close to the 
straight 
lines which are the trajectories of the linear equation. The 
trajectories to the left  of the critical line differs significally 
from the trajectories of the linear equation. In this domain the 
non-linear corrections, induces by partons interaction in the parton 
cascade,  are large and change considerably the physics of the QCD 
evolution traditionally related  to the linear evolution equation. 

\begin{figure}[htbp]
\begin{minipage}{11.0cm}
\epsfig{file= 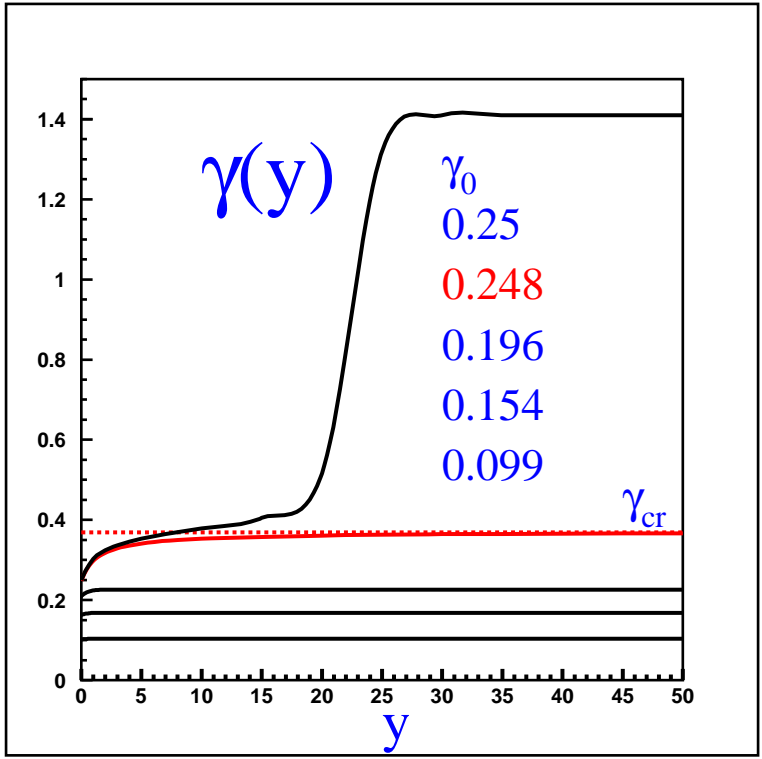,width=100mm}
\end{minipage}
\begin{minipage}{5.0 cm}
\caption{The anomalous dimension $\gamma(y)$ on the trajectories for the 
non-linear evolution equation.
}
\label{gatr}
\end{minipage}
\end{figure}

The same physical picture we can see in \fig{gatr} which shows the 
quite different behaviour of the anomalous dimension $\gamma$ along 
trajectories. For the trajectories to the right of the critical line 
$\gamma(y)$ very rapidly reaches a constant value as it should be for 
the linear evolution. On the critical line $\gamma(y)$ approaches 
$\gamma_{cr}$ (see \eq{GAMMACR} ) but very slowly. For the 
trajectories to the left of the critical line we have a different 
behaviour and $\gamma(y)$ becomes larger than unity which is, of 
course,  an indication that we cannot use a semi-classical approach, 
at least in the present form. 
 
\begin{figure}[htbp]
\begin{minipage}{11.0cm}
\epsfig{file= 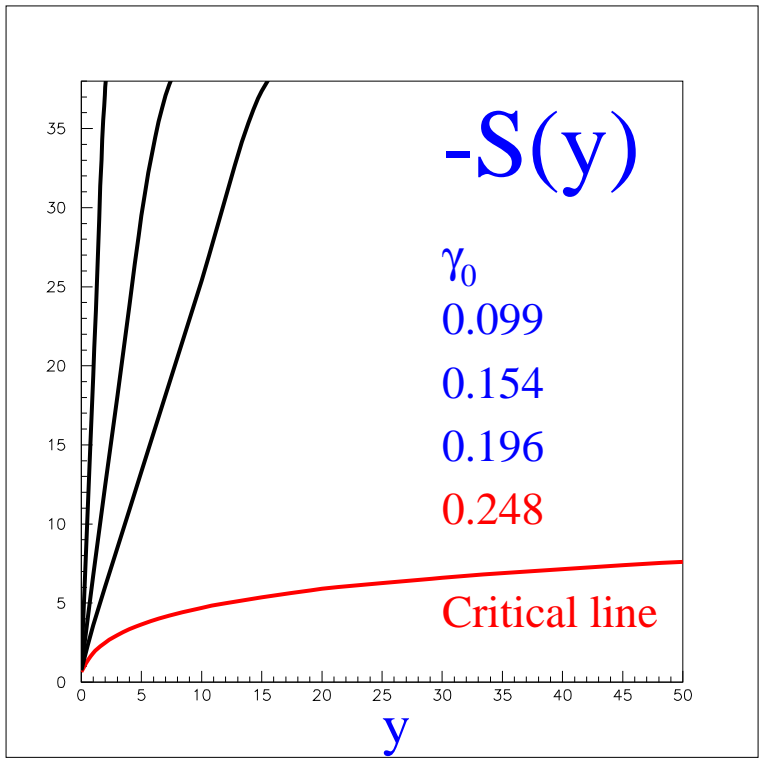,width=100mm}
\end{minipage}
\begin{minipage}{5.0 cm}
\caption{The value of $-S(y)$   on the trajectories for the
non-linear evolution equation at $b_t = 0$. The dipole amplitude is 
defined as $N
\,=\,e^{S(y)}$.}
\label{str}
\end{minipage}
\end{figure}

\fig{str} shows the behavior of function $S(y)$ ($ N = e^{S}$ ) versus $y 
= \ln(1/x)$. One can see that the dipole amplitude $N$ is decreasing 
slowly on the critical trajectory while it is rapidly falling down on the 
trajectories to the right of the critical line. It is worthwhile 
mentioning 
that at small $y$ ($ y = 0 \div 15$) the $y$- dependence stems mostly 
from the dependence of $\gamma(y)$ on $y$ ( see \fig{gatr} ). We see the 
manifestation of the slow decrease of \eq{CRITTR2}  only at 
$y\,\geq\,\,15$.

\subsubsection{A general solution}

Since trajectories cannot cross each other we see (i) that the domain to 
the 
left of the critical line does not know anything about the domain to the 
right of the critical line; and (ii) that the solution to the right of the 
critical line is very close to the solution of the linear evolution 
equation. Therefore, the semi-classical approach to the non-linear 
evolution equation leads to an  idea\cite{GLR}  to solve the problem 
considering only the 
linear evolution equation but with initial condition not at fixed value of 
$\xi = \xi_0$ but on the critical line with $N$ is defined by 
\eq{CRITTR2}. For fixed QCD coupling it is easy to find such a solution.

Indeed,  the general solution to the linear equation has a form of \eq{BFKLLB}
where $\phi_{in}$ should be defined from the initial  condition.
Using \eq{CRITTR2} and \eq{CRITTR} we obtain that 
\beq \label{SOLR1}
\phi_{in} \,\,=\,\,\,\bas\,( \frac{d \chi(\gamma)}{d \gamma} 
\,\,-\,\,\frac{d 
\chi(\gamma_{cr})}{d \gamma_{cr}}\,)\,\,e^{-( 1 - 
\gamma_{cr})\,\xi_0}\,\,e^{ 
\,-\,\frac{1}{N_0}\,\{\,\as 
\,\chi(\gamma)\,- 
\chi(\gamma_{cr})\,\frac{1\,-\,\gamma}{1\,-\,\gamma_{cr}}\,\} }\,\,.
\eeq
\eq{BFKLLB} with this $\phi_{in}$ gives the solution to the linear 
equation with the initial condition on the critical line. 

\subsubsection{Impact parameter dependence}

\eq{SCEQ1} does not explicitly depend on $b_t$. The entire $b_t$ 
dependence is hidden in initial condition (see \eq{GMBA}) both the 
initial dipole amplitude $N_0$ and the anomalous dimension $\gamma_0$ 
depend actually on the variable $\tau$ as one can see from 
\eq{GMBA}), namely,$ N_0 = N(\tau_0)$ and $\gamma_0 = \tau_0 d 
N(\tau_0)/d \tau_0$. For example, the critical line corresponds to 
$\tau_{0,cr}\,
\,=\,0.254$ and $\gamma_0 = 0.248$. 

However, when we solve \eq{SCEQ1}-1 for the trajectories, 
we need to know the initial value of $\xi$ not only of 
$\tau$. Indeed,  the 
equation for the trajectories can be solved
as
\beq \label{TREQ}
\xi\,\,=\,\,- \as\,\int^y_{y_0} 
\,\,d\,y'\,\frac{d\,\chi(\gamma(y'))}{d 
\gamma  }\,\,+\,\,\xi_0(b_t)\,\,.
\eeq
It is obvious from \eq{INCONMR} that 
\beq \label{XI0}
\xi_0(b_t)\,\,=\,\,\ln 
\bar{\tau}(y=y_0,b_t)\,\,-\,\,\ln(\tau_{0,cr}(y=y_0))\,\,.
\eeq
It should be stressed that the $b_t$ dependence also enters the 
definition of $\xi = \,\ln(k^2k'^2\,b^4_t)$. Substituting \eq{XI0} 
into the equation for the critical line (see \eq{CRITTR} ) one 
obtains the $b_t$ dependence of the saturation scale:
\beq \label{QSB}
Q^2_s(y,k'^2;b_t)\,\,=\,\,Q^2_s(y=y_0,k'^2; b_t)\,e^{\as 
\,\frac{\chi(\gamma_{cr})}{1 \,- \,\gamma_{cr}}\,( y - y_0)}
\eeq
This equation we can rewrite at large $b_t$  using the exact 
expression for 
$Q^2_s(y=y_0,k'^2; b_t)$ (see \eq{XSBA}). It has the form
\beq \label{QSLB}
Q^2_s(y,k'^2;b_t)\,\,=
\,\,\frac{1}{k'^2\,b^4_t\,\tau_{0,cr}}\,\pi\,\as^2
\frac{N^2_c - 1}{3\,N^2_c}\,(2\,
m_{\pi}\,b_t)^4\,K_4(2\,m_\pi\,b_t)\,\,\,e^{\as
\,\frac{\chi(\gamma_{cr}}{1 \,- \,\gamma_{cr}}\,( y - y_0)}
\eeq

In pQCD region \eq{QSLB} leads to 
\beq \label{QSPER}
Q^2_s(y,k'^2;b_t)\,\,=
\,\,\frac{1}{k'^2\,b^4_t\,\tau_{0,cr}}\,\pi\,\as^2
\frac{\,(N^2_c - 1)}{3\,N^2_c}\,\,\,e^{\as
\,\frac{\chi(\gamma_{cr}}{1 \,- \,\gamma_{cr}}\,( y - y_0)}\,\,.
\eeq

As we will see below this expression for $Q_s$ leads to the typical 
radius of interaction which increases as a power of energy, namely,
\beq \label{R}
R^2(y)\,\,\propto\,\,\,e^{\,\h\frac{\chi(\gamma_{cr})}{1 \,- 
\,\gamma_{cr}}\,( y - y_0)}
\eeq
in a full agreement with the analysis given in Ref. \cite{KW}.
It gives the power-like increase of the total dipole cross section, 
namely
$$ \sigma_{tot}(dipole) \,\,=\,\,2 \,\pi\,R^2\,\,\propto\,\,
e^{\,\h\frac{\chi(\gamma_{cr})}{1 \,-
\,\gamma_{cr}}\,( y - y_0)}$$
in an explicit violation of the Froissart theorem \cite{FROI}.

However, the non-perturbative contributions in the initial condition 
modify \eq{QSPER} at large values of $b_t$ and   \eq{QSLB} leads to 
$Q^2_s \,\,\propto\,\,e^{- 
2\,m_\pi\,b_t}$ for large  $b_t\,\geq\,1/2\,m_\pi$. It is worthwhile 
mentioning that the $b_t$ of the saturation scale is the only way how 
the $b_t$ dependence could affect the solution in the saturation 
domain ( to the left of the critical line in \fig{traj} ). 
We will see below that $Q^2_s \,\,\propto\,\,e^{-
2\,m_\pi\,b_t}$ generates $\sigma_{tot}(dipole)\,\propto\,\ln^2(1/x)  
$ in 
accordance with the Froissart theorem \cite{FROI}.  

\subsection{Running $\mathbf{\alpha}_{\mathbf{S}}$}

\subsubsection{A general discussion}

In this section we are looking for the semi-classical solution to our 
master equation (see \eq{EQ}) considering running QCD coupling, namely,  
$\as$ in \eq{EQ} is equal to $\as \,=\,\frac{4 \pi}{b \ln 
(1/(x^2_{01}\Lambda^2))}$. The main uncertainty in taking into account the 
running QCD coupling is the fact that we do not know what is correct way 
to incorporate running $\as$ into the linear BFKL equation 
\cite{LEREN,BRARUN,DR}. However, it was proven, see \cite{GLR}, that we can 
safely use \eq{EQ}   with running $\as$ in semi-classical approach.

It turns out that it is useful to search the semi-classical solution 
in 
the form
\beq \label{SCSOLR}
N(k,k',y;b_t)\,=\,\bas(\xi)\,e^S\,\,=\,\,\bas(\xi)\,e^{\omega(\xi,y,b_t)\,y\,
\,-\,\,(
\,1 \,-\,\gamma(\xi,y,b_t)\,)\,\xi\,\,\,
+\,\,\beta(k',b_t)}\,\,,
\eeq
where $\omega$ and $\gamma$ are smooth function of $y$ and $\xi$.

In this case the master equation has a form:
\beq \label{EQR}
\omega\,\,-\,\,\bas(\xi)\,\chi(\gamma)\,\,+\,\,\bas^2(\xi)\,e^S\,\,=\,\,0\,\,.
\eeq

Using \eq{SC2}, \eq{SCEQ} and  the explicit form of $\as$, namely
\beq \label{RAS}
\bas(\xi)\,\,=\,\,\frac{4 \,\pi}{b\,\left(\,\xi 
\,\,-\,\,\ln(k'^2\,b^4_t\,\Lambda^2)\,\right)}\,\,=\,\,\,\,\frac{4 
\,\pi}{b\,\left(\,\xi \,\,-\,\,\bar \xi(k',b_t)\right)}\,\,
\eeq
we can derive the following set of equation.

$$(1.)\,\,\,\,\,\,\,\frac{d\,\xi}{d\,t}\,\,=\,\,F_\gamma \,\,\,=\,\, 
-\,\bas(\xi)
\,\frac{d\,\chi(\gamma)}{d\ \gamma}
\,\,;\,\,\,\,\,\,\,\,\,\,\,\,(2.)\,\,\,\,\,\,\,\,
\frac{d\,y}{d\,t}\,\,=\,\,F_\omega\,\,=\,\,1\,\,;$$
$$(3.)\,\,\,\,\,\,\,\,
\frac{d\,S}{d\,t}\,\,=\,\,\gamma\,F_\gamma \,+\,\omega\,F_\omega\,\,=\,\,
\bas(\xi)\,(1 - \gamma)\,\frac{d\,\chi(\gamma)}{d\,\gamma}
\,+\,\omega\,\,;$$
$$ (4.)\,\,\,\,\,\,\,\,
\frac{d\,\gamma}{d\,t}\,\,=\,\,-\,(F_\xi\,+\,\gamma\,F_S)\,\,=\bas^2(\xi)\,
(\,1\,-\,
\gamma\,)\,e^{S}\,\,-\,\,\bas^2(\xi)\,\frac{b}{4\,\pi}\,\chi(\gamma)\,\,+
\,\,\bas^3(\xi)\,\frac{b}{2 
\pi}\,e^S\,\,;$$
\beq \label{EQR1}
(5.)\,\,\,\,\,\,\,
\frac{d\,\omega}{d\,t}\,\,=\,\,-\,(F_y\,+\,\omega\,F_S)\,\,=\,\,-
\,\,\bas^2(\xi)\,\omega\,e^{S}\,\,;
\eeq

\subsubsection{Critical trajectory}

There exists a critical trajectory in this case as well as in the 
case of fixed $\as$. To see it let us assume that $e^S \,\approx \,1$ 
and $\as \,\ll\,1$. For such $S$ we can neglect the non-linear term 
in \eq{EQR} and \eq{EQR1}-3 has a form:
\beq \label{EQR2}
\frac{d S}{d y}\,\,=\,\,\bas \,\left(\, \chi(\gamma) \,+\,( 
1\,-\,\gamma)\,\frac{d \chi(\gamma)}{d \gamma}\,\right)\,\,.
\eeq

One sees from \eq{EQR2} that there is a trajectory with the same 
equation for the anomalous dimension ($\gamma \,=\,\gamma_{cr}$)  as 
in 
the case of constant $\as$ (see \eq{GAMMACR}) on which $S$ is 
constant. \eq{EQR1}-4 allows us to determine this constant. Indeed, 
we can see from \eq{EQR1}-4 that $\gamma$ on this trajectory would be 
constant if \cite{GLR}
\beq \label{SCR} 
e^{S_{cr}}\,\,=\,\,\frac{b}{4\,\pi}\,\frac{\chi(\gamma_{cr})}
{1\,-\,\gamma_{cr}}
\eeq

The form of this critical trajectory is given by \eq{EQR1}-1, namely
\beq \label{CRTREQ} 
\frac{d \xi_{cr}(y)}{d y}\,=\,- \bas(\xi)\,\frac{d 
\chi(\gamma_{cr})}{d \gamma_{cr}}\,\,,
\eeq
which leads to

\beq \label{CRTREQ1}
(\,\xi \,\,-\,\,\bar \xi(k',b_t)\,)^2\,\,-\,\,(\,\xi_0(b_t) 
\,\,-\,\,\bar 
{\xi}(k',b_t)\,)^2\,\,=\,\,\frac{8\,\pi}{b}\,
\mid\frac{d \chi(\gamma_{cr})}{d \gamma_{cr}}\mid\,(y\,\,-\,\,y_0\,)\,\,,
\eeq
where $\xi_0(b_t)$ and $y_0$ are determined by the initial 
conditions.

Therefore, we have constant $S_{cr}$ and  $\gamma_{cr}$ on the 
critical trajectory of \eq{CRTREQ1}. One can see from 
 \eq{EQR1}-5
that $\omega$ 
is  
also approximately constant on this line , namely, 
\beq \label{CRTROM}
\omega\,\,\propto\,\,e^{- \frac{4\,\pi}{b}\,\frac{1}{|\frac{d 
\chi(\gamma_{cr})}{d 
\gamma_{cr}}|}\,\ln(\xi)}\,\,.
\eeq 

\subsubsection{$\mathbf{b}_{\mathbf{t}}$-dependence of the saturation 
scale}

\begin{figure}[htbp]
\begin{minipage}{11.0cm}
\epsfig{file= 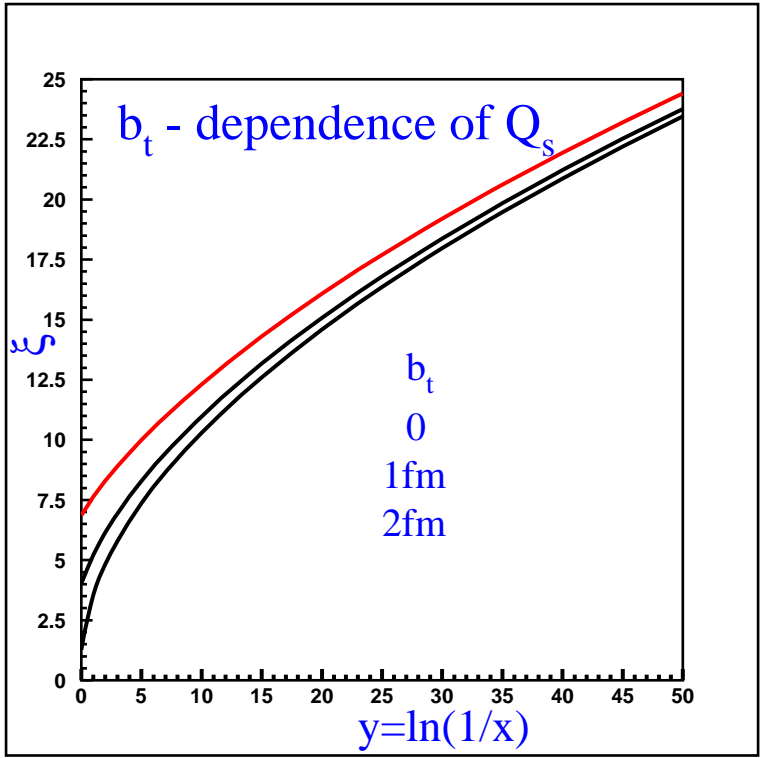,width=100mm}
\end{minipage}
\begin{minipage}{5.0 cm}
\caption{The behavior of the saturation scale as function of $y$ at 
different values of the impact parameters ($b_t$). As one can see at 
large values  of $y$ the saturation scale does not depend on $b_t$ in 
a striking difference with the case of frozen QCD coupling.} 
\label{btqs}
 \end{minipage}
 \end{figure}

The saturation scale, which is the value of the typical virtuality 
$k^2$ on the critical line, can be determined resolving \eq{CRTREQ1} 
with respect to $k^2$. It is easy to see that
\beq \label{SATSCR}
Q^2_s(y,b_t)\,\,=\,\,\Lambda^2\,e^{\sqrt{(\,\xi_0(b_t)\,\,-\,\,
 \bar{\xi}(k',b_t)\,)^2 \,\,\,+\,\,\frac{8\,\pi}{b} 
|\chi'_{\gamma_{cr}}(\gamma_{cr})|\,(y\,-\,y_0\,)}}
\,\,,
\eeq
where 
$$\xi_0(b_t)\,-\,\bar{\xi}(k',b_t)\,\,=\,\,\ln \left(
\frac{k'^2m^4_\pi\,K_4(2\,m_\pi\,b_t)}{\,\tau_{0,cr}\,}\right)\,
\,\longrightarrow|_{b_t\,\gg\,1/2 m_\pi}\,\,- 2 
m_\pi\,b_t \,\,,
$$
for the initial conditions given by \eq{GMBA}.
However, we  will discuss below that we cannot solve the non-linear 
evolution equation with running QCD coupling with this initial 
condition. We need to evolve first the dipole amplitude with the 
linear BFKL equation to some value of $y =y_{in}=\ln(1/x_{in})$ and 
only at 
$y=y_{in}$ we can use the expression of \eq{GMBA} for the initial 
condition. It leads to substitution $N^{BFKL}(x,y=y_{in};b_t)$ 
instead 
of  $N^{BA}(x,y=y_0;b_t)$ 
in 
\eq{GMBA}. For the solution to the BFKL equation we use \eq{BFKLLB}
which give us the following equation for  $N^{BFKL}(x,y=y_{in};b_t)$ 
:
\beq \label{INCONBFKL}
N^{BFKL}(x,y=y_{in};b_t)\,\,\,=\,\,\,\h\,\tilde{\tau}\,
e^{\omega_L\,(y_{in} 
- y_0)}\,\,e^{\h \xi}\,\,.
\eeq
$e^{\h \xi}$ reflects the fact that the BFKL anomalous dimension is 
equal to $\h$ and $\omega_L = \bas \chi(\h)$ \cite{BFKL}. 
\eq{INCONBFKL} leads to the following expression for 
the saturation scale at $y = y_{in}$:
\beq \label{QSYIN}
Q^2_s(y=y_{in},k'^2;b_t)\,\,=
\,\,(4\,\tilde{\tau}\,)^2\,e^{2\,\omega_L\,(y_{in}
- y_0)}\,k'^2\,(\tau_{0,cr}\,(\,k'^2\,b^2_t\,+\,1))^{-2}\,\,
\eeq
$$
\longrightarrow|_{b_t \,>\,1/k'} 
\,\,\,\frac{1}{k'^2\,b^4_t\tau_{0,cr}^{2}}\,\,\left(\,(2
m_{\pi}\,b_t)^4\,K_4(2\,m_\pi\,b_t)\,\,
e^{\,\omega_L\,(y_{in}- y_0)}\right)^2\,
$$

\eq{QSYIN} translates in
$$\xi_0(b_t)\,-\,\bar{\xi}(k',b_t)\,\,=$$

\beq \label{DIF}  
\,\,\ln(Q^2_s(y=y_{in},k'^2;b_t)/\Lambda^2)\,=\,\left \{ 
\begin{array}{l} - \ln(b^4_t\,k'^2 \Lambda^2) 
\,\,\mbox{for}\,\,(1/2\,m_\pi)\,\gg\,b_t\,\gg\,1/k' \\ \\ \\
-  4 m_\pi\,b_t \,\,\mbox{for}\,\,b_t\,\gg\, (1/2\,m_\pi)\
\end{array} \right.
 \eeq
   
\eq{SATSCR} shows that the saturation scale does not depend on $b_t$ 
for $b_t \,\leq\,b_0(y)$ with $b_0$ which is the solution of the 
following equation
\beq \label{B0}
ln(Q^2_s(y=y_{in},k'^2;b_0)/k^2)\,\,=\,\,\sqrt{\frac{8\,\pi}{b} 
|\chi'_{\gamma_{cr}}|\,(\gamma_{cr})\,(y\,-\,y_0\,)}.
\eeq
From \eq{DIF} one can see that $b_0\,\propto\,exp\left(\h 
\sqrt{\frac{8\,\pi}{b}
|\chi'_{\gamma_{cr}}(\gamma_{cr})|\,(y\,-\,y_0\,)} \right)$ at $1/2 
\,m_\pi \,\gg\,b_0\,\gg\,1/k'$ and  $b_0 \,\,=\,\,(1/4 
m_\pi)\,\sqrt{\frac{8\,\pi}{b}
|\chi'_{\gamma_{cr}}(\gamma_{cr})|\,(y\,-\,y_0\,)}$ for $b_0 \,\gg\,1/2 
m_\pi$. In other words, $Q^2_s \,\, \rightarrow\,\,1/b^4_t$ for $b_t 
\,>\,b_0 $ if $ 1/k'\,<\,b_0\,<\,1/2\,m_\pi$ 
and the saturation scale  $Q^2_s\,\,
\rightarrow\,\,e^{- 4\,m_\pi 
\,b_t} $ for $b_0 \,>\,1/2\,m_\pi$.
\fig{btqs} illustrates such a behaviour of the saturation scale.
It should be stressed that we have a different large $b_t$ behaviour 
for the running $\as$ due to different initial conditions.

\subsubsection{Numerical solution}

\begin{figure}[htbp]
\begin{minipage}{11.0cm}
\epsfig{file= 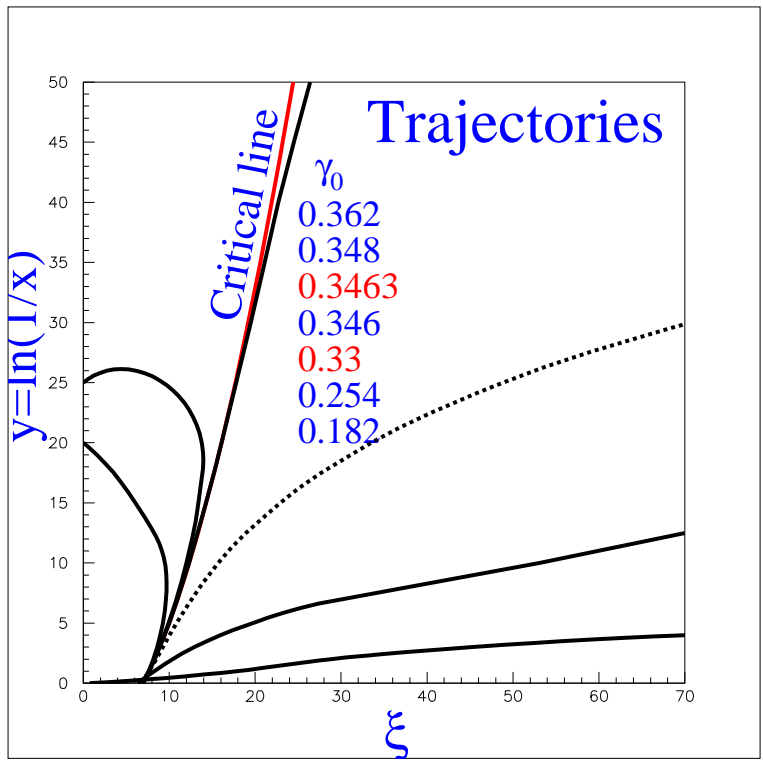,width=100mm}
\end{minipage}
\begin{minipage}{5.0 cm}
\caption{Trajectories for the non-linear evolution equation ( see 
\eq{EQR1})
in semi-classical approach
at $b_t = 0$  for running QCD coupling.he dotted line corresponds to 
the trajectory which is tangent to the critical line.}
\label{trajra}
\end{minipage}
\end{figure}

\begin{figure}[htbp]
\begin{minipage}{11.0cm}
\epsfig{file= 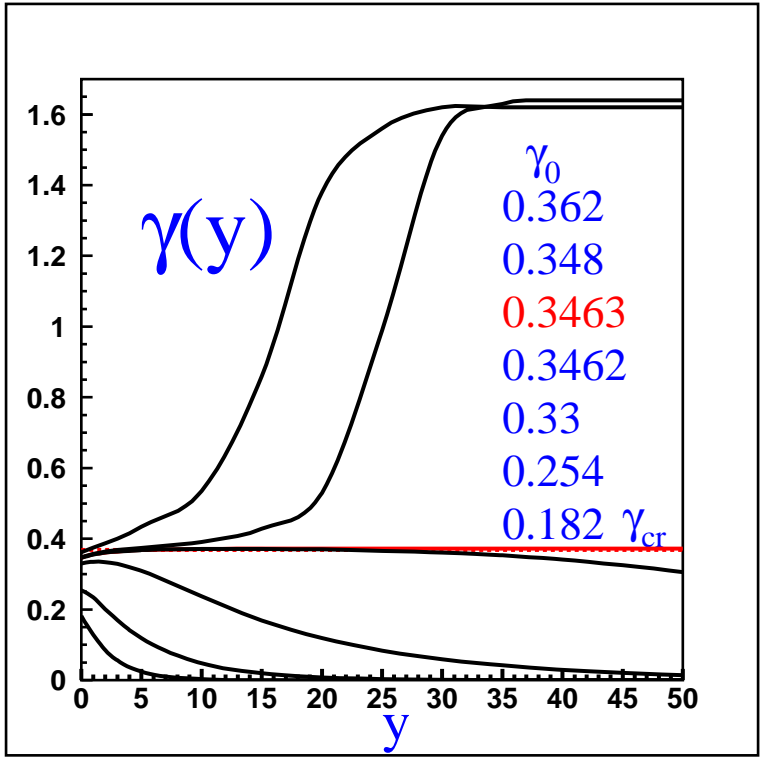,width=100mm}
\end{minipage}
\begin{minipage}{5.0 cm}
\caption{The anomalous dimension $\gamma(y)$ on the trajectories for the
non-linear evolution equation (see \eq{EQR1}).
The dotted line corresponds to $\gamma= \gamma_{cr}$.}
\label{gatra}
\end{minipage}
\end{figure}

\begin{figure}[htbp]
\begin{minipage}{11.0cm}
\epsfig{file= 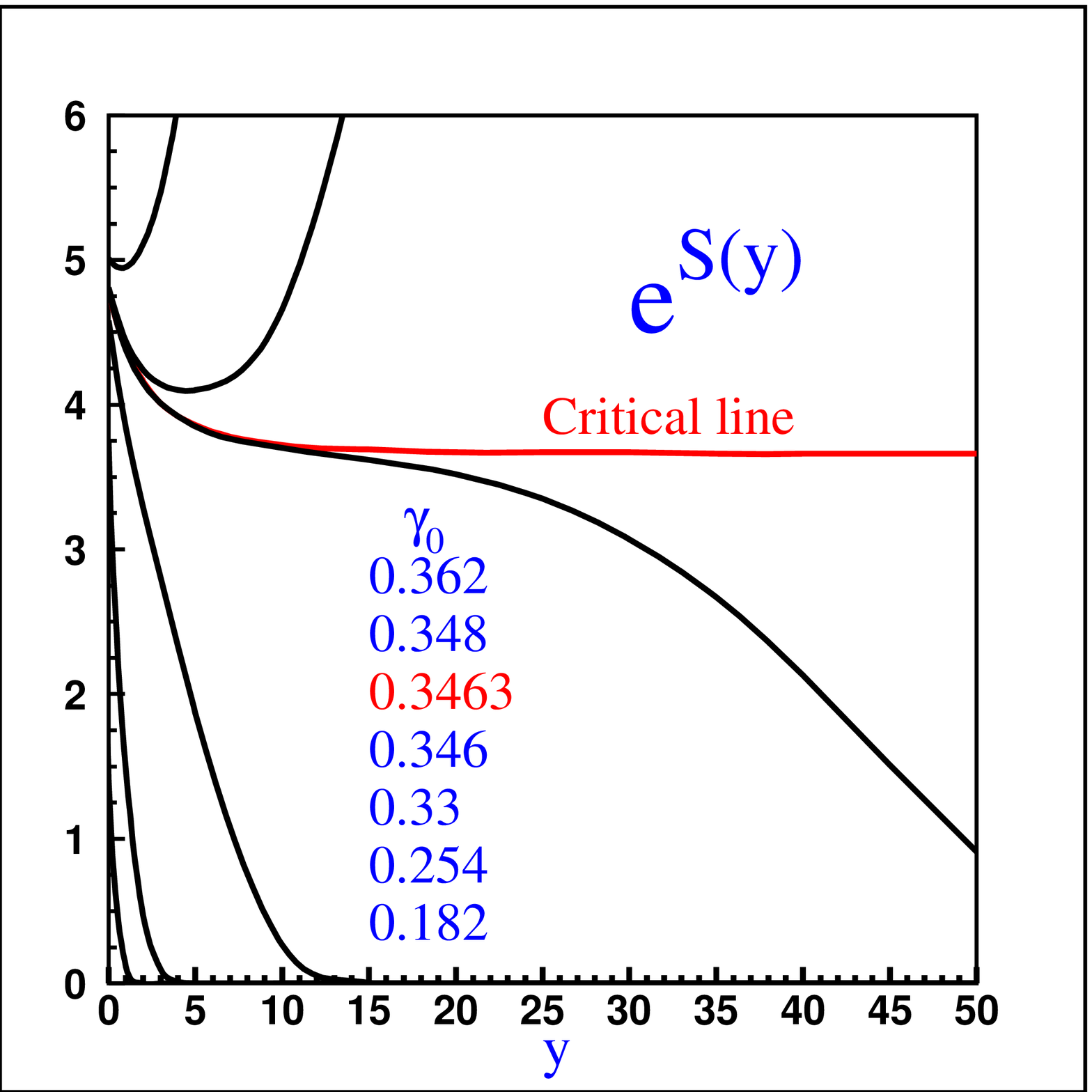,width=100mm}
\end{minipage}
\begin{minipage}{5.0 cm}
\caption{The value of $e^{S(y)}$ on the trajectories of the non-linear 
evolution equation   (see \eq{EQR1}) at $b_t=0$. The dipole amplitude is 
defined as $N = \as \,e^{S(y)}$. }
\label{nra}
\end{minipage}
\end{figure}

Our above discussion we support by numerical solution of \eq{EQR1}. We 
simplify a bit this  set of equations neglecting $\bas^2$ contribution in 
\eq{EQR} and $\bas^3$ term in \eq{EQR1}-4.
\fig{trajra} shows that the trajectories for the running QCD coupling have 
the same pattern as for frozen $\as$: 
\begin{itemize}
\item \quad They can be divided in two parts by the critical line which 
has the form of \eq{CRTREQ1} starting from $\xi \,\geq 5$. The anomalous 
dimension on this line rapidly reaches the value of $\gamma_{cr}$ ( see 
\eq{GAMMACR} and \fig{gatra} )  while $S(y)$ is constant at large $y$ (see 
\fig{nra});
\item \quad  All trajectories to the left of critical line in \fig{trajra}
go to the negative values of $\xi$ with unreasonable values of the 
anomalous dimension $\gamma(y) \,\rightarrow \,1.62$ (see \fig{gatra})  
and with very fast 
increase of $e^{S(y)}$ shown in \fig{nra}. Such a behaviour reflects the 
fact that we cannot apply the semi-classical method for the solution 
inside of the saturation (CGC) region at least when the trajectories 
go far away from the critical line. Comparing \fig{gatr} and \fig{traj} 
with \fig{gatra} and \fig{trajra} we conclude that solutions for both  
frozen  and running QCD coupling have the same qualitative features;
\item \quad  Among  all trajectories to the right of the critical line we  
can 
conditionally separate two groups. The trajectories of the first group
envelope the critical line and only at some value of $y$ they go apart. 
For the large values of $y \,\geq\,y_{cr}$ they coincide with the 
trajectories of the 
linear evolution equation. Therefore we can find the dipole amplitude on 
such trajectories just solving the linear evolution equation with the 
initial condition $N = \bas e^{S_{cr}}$ at $y =y_{cr}$ with $e^{S_{cr}}$ 
given by \eq{SCR}.
 The trajectories which do not touch the 
critical line belong to the second group. They are the same as 
trajectories for the linear equation. 
\item \quad The trajectory which separate these two groups of 
trajectories to the right of critical line is denoted by dotted line 
in \fig{trajra} and it touches the critical line in the only one 
point. The equation of this tangent  line was found in Ref.\cite{GLR}
and it reads:
\beq \label{TANG}
y -\hat{y}(\xi,y)\,\,=\,\,\h\,\left( \frac{4 \pi}{b \as 
(\hat{\xi}(\xi,y))}\right)^2\,\,\ln \xi\,\,,
\eeq
where $\hat{y}(\xi,y)$ and $\hat{\xi}(\xi,y))$ are the coordinates of 
the point where the tangent line ( dotted line in \fig{trajra}), 
passing through the point $(\xi,y)$, touches the critical line. 
 \end{itemize}

The essential difference between frozen and running QCD coupling occurs 
in the initial condition for the non-linear equation. For the frozen $\as$ 
we used the initial conditions given by \eq{GMBA} and \eq{XSBA}. However, 
we  found out the the non-linear equation with running $\as$ cannot be 
solved with the same initial condition. It turns out that we need to 
evolve our parton cascade using the BFKL equation (see \eq{BFKLLB}) for 
$y$ 
less than $y_0$  ($x \,\geq\,x_0$) and use \eq{GMBA} as the initial 
condition at $y = y_0 $ ($ x = x_0$) with $N^{BFKL}$ instead $N^{BA}$.
In our numerical calculation we use \eq{BFKLLB} and \eq{INPHI} with $x_0= 
10^{-1}$ to obtain the solution that has been discussed.

\section{Semi-classical Approach in Saturation  (Colour Glass 
Condensate ) Domain}

\subsection{Reduction of the non-linear equation to the 
semi-classical form}

As have been discussed  we cannot expect to solve the master equation 
in 
the saturation (CGC) domain considering the dipole  amplitude
$\tilde{N} = e^S$ such as $S$  being smooth functions of $y$ and 
$\xi$.  Therefore, the fist question, which we have to answer, is for
what function we are going to use the semi-classical approximation.
We  follow the idea of Ref. \cite{LT} and introduce the new function 
$\Phi$ as
\beq \label{PHI}
\tilde{N}(y,\xi;b_t)\,\,=\,\,\frac{1}{2}\,\int_{\xi}\,\,d \xi'\,
\left(\,1\,\,-\,\,e^{- \Phi(y,\xi';b_t)}\,\right)\,\,.
\eeq

We consider $\Phi $ as a smooth function such that 
$\Phi_{\xi \xi}\,\,\ll\,\,\Phi_\xi\,\Phi_\xi$, $ \Phi_{y 
y}\,\,\ll\,\,\Phi_y\,\Phi_y $ as well as $\Phi_{\xi y}
\,\,\ll\,\,\Phi_\xi\,\Phi_y$. Therefore,
\beq \label{DERPHI}
\frac{\partial^n}{(\partial \xi)^n}\,e^{- \Phi}\,\,=\,\,(-\Phi_{\xi})^n
\,e^{- \Phi}\,\,.
\eeq
Substituting \eq{DERPHI} in \eq{EQK} we reduce it to the following 
equation for  function $\Phi$:
\beq \label{EQPHI}
-\,\frac{\partial^2 \Phi(y,\xi;b_t)}{\partial y\,\partial 
\xi}\,\,=\,\,\bas\,\left(\,1\,\,-\,\,e^{- 
\Phi(y,\xi;b_t)}\,\right)\,\,+\,\,\bas\,\frac{d L(-\Phi_\xi)}{d 
\Phi_\xi}\,\frac{\partial^2 
\Phi(y,\xi;b_t)}{\partial \xi\,\partial
\xi}\,,
\eeq
where $L(f)\,=\,(f\,\chi(1 - f)\,-\,1)/f$. In double log approximation of 
pQCD which was discussed in Ref.\cite{LT} function $L$ is equal to zero 
since $\chi(1 - f) \,=\,1/f$. For the derivation of \eq{EQPHI} one needs 
to take into account that the series for $L(f)$ starts with $f^2$ term.

In this section we will solve \eq{EQPHI} using the powerful idea of the 
geometrical scaling behaviour which will allow us to simplify this 
equation.

\subsection{Geometrical scaling}

Inside of the saturation domain we expect so called geometrical 
scaling behaviour of the dipole amplitude, namely, this amplitude 
depends on one variable $\tau \,=\,Q^2_s(x;b_t)/k^2$ instead of 
three: $x$, $k^2$ and $b_t$. The physics of such behaviour has been 
discussed for long time starting from the first papers where the 
saturation scale was introduced \cite{GLR,MUQI,MV}. Recently, it has 
been 
found that the HERA data supports this idea in the wide range of low 
$x$ ( $x\,\leq\,10^{-1}$) (see Ref. \cite{KSC} for details) as well 
as the numerical solutions of the non-linear evolution equation show 
this scaling behaviour \cite{LLU,AB,GMK}. However, the theoretical 
proof \cite{LT,IIM,BALE}   of this scaling behaviour is still in a 
very embryonic state.
The only result, that has been proven so far, is that in the 
semi-classical 
approach the master non-linear equation has a geometrical scale 
solution both in the saturation (CGC)   region \cite{BALE} as well as 
for \cite{IIM,GLR}\footnote{It is interesting to mention that the 
semi-classical solution in  Ref. \cite{GLR} shows the geometrical 
scale behaviour, but it was found out only after an explanation given 
in Ref. \cite{IIM}.} $ 
(Q^4_s/\Lambda^2)\, >\,Q^2>\,  Q^2_s $.

In this paper we will find the semi-classical solution to 
\eq{EQK}   inside 
of the 
saturation (CGC) region. 

We start with the general arguments why the semi-classical leads to 
the geometrical scale behaviour of the dipole amplitude given in Ref. 
\cite{BALE}.

The solution to the linear equation can be written through the 
moments of the dipole amplitude, namely,
\beq   \label{MOM}
\tilde{N}(k,k',y;b_t)\,\,=\,\,\int\,\,\frac{d
\omega}{2\,\pi\,i} \,\phi(k',\omega;
b_t)\,e^{\omega \,y\,\,- \,\,( 1\,-\,\gamma(\omega))\,\xi}\,
\eeq
where $\gamma(\omega)$ is a solution to the following equation:
\beq \label{BFKLGA}
\bas\,\,\,\chi(\gamma(\omega)) \,\,=\,\,\omega\,.
\eeq

In $\omega$ representation \eq{EQK} has a form
\beq \label{EQOM}
\left(\,\omega\,\,-\,\,\bas 
\chi(\gamma(\omega))\,\right)\,N(\omega,\xi;b_t)\\,\,=\,\,- 
\,\,\bas \,\,\int \frac{d 
\omega'}{2\,\pi\,i}\,N(\omega',\xi;b_t)\,N(\omega 
\,-\,\omega',\xi;b_t)\,.
\eeq

In the semi-classical approach we can use the steepest decent method 
to take the integral over $\omega'$ in \eq{EQOM} and the saddle point 
in this integration is equal to $\omega'_S= \omega/2$. After doing 
this we obtain:
\begin{eqnarray} 
\left(\,\omega\,\,-\,\,\bas
\chi(\gamma(\omega))\,\right)\phi(k',\omega;b_t)\,e^{-( 
1\,-\,\gamma(\omega))\,\xi}\,\,&=&      \label{EQSDP} \\
 &-&\,\,\bas 
\phi(k',\frac{\omega}{2};b_t)\,\frac{1}{2\,\sqrt{2\pi\,
\gamma_{\omega\omega}\,\xi}}\,e^{\,-2 
(1\,-\,\gamma(\frac{\omega}{2}))\,\xi} \nonumber
\end{eqnarray}
In \eq{EQSDP} one can see two different regions in $\omega$: 
\begin{enumerate}
\item \quad $ \omega\,\,\gg\,\,\omega_{crit}$. Here we can neglect the 
r.h.s. in \eq{EQSDP} and anomalous dimension $\gamma(\omega)$ can be 
calculated from \eq{BFKLGA}.
\item \quad $ \omega\,\,\ll\,\,\omega_{crit}$.  The r.h.s. of \eq{EQSDP}
is not small and at the fist requirement for the solution we need to have 
\beq \label{ANDIMSC}
( 1 \, -\, \gamma(\omega)|_{\omega\,\,\ll\,\,\omega_{crit}})\,\,=\,\,2\,( 
1 \, -\, \gamma(\frac{\omega}{2})|_{\omega\,\,\ll\,\,\omega_{crit}})\ 
\,\,.
\eeq
This equation has a simple solution \cite{BALE}
\beq \label{ANDIMSCSOL} 
( 1 \, -\, 
\gamma(\omega)|_{\omega\,\,\ll\,\,\omega_{crit}})\,\,=\,\,C\,\omega\,\,.
\eeq
\item \quad
Both solutions should match at a certain value  $ 
\omega\,\,=\,\,\omega_{crit}$.
Therefore, we have
\begin{eqnarray}
( 1\, -\, \gamma(\omega))|_{\omega\,\,\gg\,\,\omega_{crit}}\,&=&\,
(1\,-\,\gamma(\omega))|_{\omega\,\,\ll\,\,\omega_{crit}}\,\,;\label{MAT1}\\
\frac{d \gamma(\omega)}{d 
\omega}|_{\omega\,\,\gg\,\,\omega_{crit}}\,&=&\,
\frac{\gamma(\omega)}{d 
\omega}|_{\omega\,\,\ll\,\,\omega_{crit}}\,\,;\label{MAT2}\\ 
-\,\frac{\gamma(\omega)}{d
\omega}|_{\omega\,\,\ll\,\,\omega_{crit}}\,&=&\frac{ 1 - \gamma(\omega)}{d 
\omega}|_{\omega\,\,\ll\,\,\omega_{crit}}\,\,;\label{MAT3}\\
C \,&=&\,\frac{1 - \gamma_{cr}}{\bas\,\chi(\gamma_{cr})}\,\,.\label{MAT4}
\end{eqnarray}
One can see that \eq{MAT3} is the same as \eq{GAMMACR} with 
$\gamma(\omega_{crit}) \,\equiv\,\gamma_{cr}$.
\end{enumerate}
Of course, we can see that $( 1 - \gamma ) = C \omega$ directly from 
\eq{SCEQ1} but we consider the above discussion (see Ref. \cite{BALE} for 
more details) as more general since we use the very general properties 
of the semi-classical approach but not the method of characteristics.

The fact that $( 1 - \gamma \propto \omega$ in the saturation(CGC) domain 
means that the dipole amplitude is actually a function of one variable
\beq \label{Z}
z\,\,=\,\,\bas\,\frac{\chi(\gamma_{cr})}{( 1 - 
\gamma_{cr})}\,y\,\,-\,\,\xi\,\,+\,\,\beta(k',b_t)\,\,.
\eeq

One can derive this directly from \eq{MOM} substituting \eq{MAT4} and 
changing the integration variable from $\omega$ to $\omega' = \bas 
\,\chi(\gamma_{cr})\,\omega$.

\subsection{Solution}

Looking for the function of the new variable $z$ given by \eq{Z} we 
reduce  \eq{PHI} and \eq{EQPHI} to the form:
\beq \label{SRSOL1}
\tilde{N}(z)\,\,=\,\,\frac{1}{2}\,\int^{z}_{z_0}\,\,d z'\,
\left(\,1\,\,-\,\,e^{- \Phi(z')}\,\right)\,\,;
\eeq
\beq \label{SRSOL2}
\bas\,\frac{\chi(\gamma_{cr})}{1 \,-\,\gamma_{cr}}\,\,\frac{d^2\, 
\Phi(z)}{(d 
\,z)^2}\,\,=\,\,\bas\,\left(\,1\,\,-\,\,e^{-
\Phi(z)}\,\right)\,\,-\,\,\bas\,\frac{d L(\Phi_z)}{d
\Phi_z}\,\frac{d^2\,
\Phi(z)}{( d\,z)^2}\,\,\,.
\eeq

We can  rewrite \eq{SRSOL2} as a set of equations:
\begin{eqnarray}
\frac{d\,\Phi(z)}{d z} \,&=&\,D(z)\,\,;\label{SRSOL3}\\
\left( \frac{\chi(\gamma_{cr})}{1 
\,-\,\gamma_{cr}}\,\,+\,\,\frac{d\,L(D(\Phi))}{d 
D(\Phi)}\,\right)\,D(\Phi)\,\frac{d\,D(\Phi)}{d \Phi}\,&=&
\,\,1\,\,\,-\,\,e^{- \Phi}\,\,;\label{SRSOL4} 
\end{eqnarray}
which can be easily solved. It should be stressed that \eq{SRSOL4} 
does not depend on parameters and, therefore, it could be solved 
numerically without loosing any generality.

Integrating \eq{SRSOL4} over $D$ and $\Phi$ and introducing a new 
variable $\tilde{\gamma}_{cr} = 1 - \gamma_{cr}$ we obtain:
\beq \label{SRSOL5} 
\frac{\chi(\tilde{\gamma}_{cr})}{\tilde{\gamma}_{cr}}\,\frac{D^2 - 
D^2_0}{2}\,\,+\,\,\chi(D)\,D \,-\, 
\chi(\tilde{\gamma}_{cr})\,\tilde{\gamma}_{cr}\,-\,
\int^D_{\tilde{\gamma}_{cr}}\,\chi(D')\,d D' 
\,+\,\ln\frac{D}{D_0}\,\,=
\eeq
$$
\,\,\Phi\, +\,e^{-\Phi}\,-\,\Phi_0  
-\,e^{-\Phi_0}\,\,,
$$
where $D_0=D(z_0)$ and $\Phi_0=\Phi(z_0)$ are initial values for 
$D(z)$ and $\Phi(z)$. 

We can see two distinct regions of $\Phi$ where we can obtain an 
analytic solutions to \eq{SRSOL3} and \eq{SRSOL4}.
\begin{itemize}
\item \quad  $\mathbf{z}\,\,\mathbf{\gg}\,\,\mathbf{z}_{\mathbf{0}}$ 
where $z_0$ 
is the value of $z$ for 
the critical line as has been discussed. In this case $D$ is 
approaching the singularity $D = 1$ in the r.h.s. of \eq{SRSOL5}. 
Therefore this equation is reduced to
\beq \label{SRSOL6} 
\frac{1}{ 1 -D}\,\,\,=\,\,\Phi\,\,,
\eeq
or $ D\, = \,1\, -\, 1/\Phi$.
Solving \eq{SRSOL3} we obtain
$\Phi \,=\,z\,-\,\ln z$. Therefore, \eq{SRSOL6} leads to the 
following dipole amplitude for large $z$
\beq \label{SRSOL7} 
\tilde{N}(z)\,=\,\h \int^z_{z_0} 
\,\left(\,1\,-\,z'\,e^{-z'}\,\right)\,d\,z'\,\,.
\eeq
Therefore, we see that at large values of $z$ our function $\Phi$ is 
also large and we can safely use the semi-classical approach.

\item \quad   
$\mathbf{z}\,\,\mathbf{\rightarrow}\,\,\mathbf{z}_{\mathbf{0}}$ . For 
such $z$ we cannot use the  semi-classical approach for function 
$\Phi$. As we have discussed we can use here rather the 
semi-classical approach for the dipole amplitude $\tilde{N}$, which 
is equal to
\beq \label{NCR}
\tilde{N}(z)|_{z \rightarrow z_0}\,\,
=\,\,N(z_0)e^{\tilde{\gamma}_{cr}\,z}\,\,
\eeq
near to the critical line.
 
However, if $\Phi_0=\Phi(z_0)$ turns out to be so  small that we can 
expand \eq{PHI} we can obtain from \eq{NCR} that 
\beq \label{PHI0}
\Phi(z)|_{z \rightarrow 
z_0}\,\,\,=\,\,\,2\,\tilde{\gamma}_{cr}\,N(z_0)\,
e^{\tilde{\gamma}_{cr}\,z}  
\eeq
In this region we can  re-derive  \eq{EQPHI} replacing $\Phi_{\xi}$ by 
$( \ln \Phi)'_{\xi}$. 
\end{itemize}

\begin{figure}[htbp]
\begin{minipage}{11.0cm}
\epsfig{file= 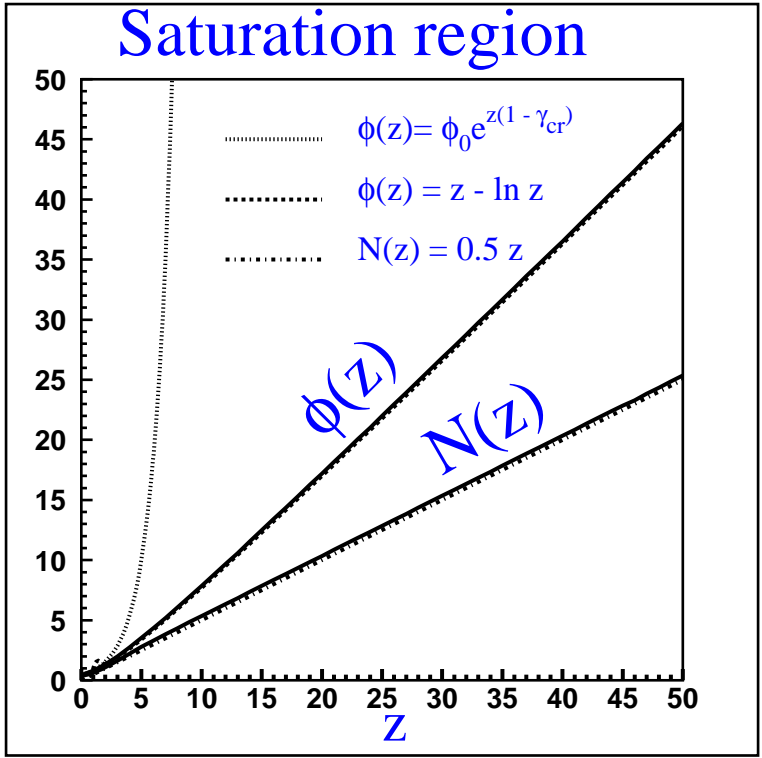,width=100mm}
\end{minipage}
\begin{minipage}{5.0 cm}
\caption{The semi-classical solution to the non-linear equation as a 
function of the geometrical scaling variable 
$z\,\,=\,\,\ln(Q^2_s(x;b_t)/k^2)$. }
\label{scphi}
\end{minipage}
\end{figure}

In \fig{scphi} we presented the numerical solution to \eq{SRSOL3} and 
\eq{SRSOL4}. We use $N_0 = 0.32$ as the  initial condition which 
comes from the solution to the right of the critical line for running 
QCD coupling. 

One can see that  the exact solution can be well approximated by the 
asymptotic formula $\Phi = z \,-\,\ln(z)$ starting from $z   = 1$.
It should be stressed that this formula leads to  a quite different 
asymptotic  behavior than it occurs in the so called double log 
approximation of pQCD (DLA) (see Ref. \cite{LT}). In DLA it turns out that 
$\Phi \,\,\propto\,\,z^2 $ deeply in the saturation (CGC) region.
 For 
lower value of $z$ we can use \eq{PHI0} as seen from  \fig{scphi} .

\subsection{Saturation for running $\mathbf{\alpha}_{\mathbf{s}}$}

As has been discussed it is not theoretically clear how to include the 
running QCD coupling in the non-linear evolution equation (see \eq{EQ}). 
In Ref. 
\cite{GLR} it was shown that we can safely consider $\as$ as a function of 
the initial dipole $x_{01}$ for the kinematic region to the right of the 
critical line. However inside the saturation region we certainly cannot do 
this. We decided to consider the running $\as$ in \eq{EQ} to be frozen on 
the critical line \footnote{In other words we take
$\,\bas\,=\,\bas(Q^2_s(y))$ inside of the saturation region.}. 
We cannot prove this assumption but it looks natural 
from the point of view that physics of saturation is determined by one 
scale: the saturation momentum.

In framework of this approach we obtain the same set of equations (see 
\eq{SRSOL3} and \eq{SRSOL4} ) as for fixed $as$. The only difference is 
the new definition of the variable $z$. Instead of \eq{Z} we have
\beq \label{ZR}
z\,\,=\,\,2\,\bas(Q^2_s)\,\frac{\chi(\gamma_{cr})}{( 1 -
\gamma_{cr})}\,y\,\,-\,\,\xi\,\,+\,\,\beta(k',b_t)\,\,.
\eeq
Using \eq{SATSCR} we can rewrite \eq{ZR} in the form:
\beq \label{ZR1}
z\,\,=\,\,\ln(Q^2_s(y;b_t)/k^2)\,\,.
\eeq
However, the statement that \eq{ZR1} as well as the fact that we have the 
same set of equations, is correct only for high 
energy ( large values of $y$ ) for which
$$\frac{8\,\pi}{b}
|\chi'_{\gamma_{cr}}(\gamma_{cr})|\,(y\,-\,y_0\,)\,\,\gg\,\,
(\xi_0(b_t)\,\,-\,\,\bar{\xi}(k',b_t)\,)^2 . 
$$

\section{Unitarity Bound}

The solution of the previous section allows us to calculate a high energy 
behavior of the observables.
     
{\bf xG($\mathbf{Q}^\mathbf{2}$,x)}. The dipole amplitude in momentum 
representation,  
that has been  calculated, is 
closely related to the gluon structure function, namely:
\beq \label{GNRE}
xG(Q^2,x)\,\,=\,\,\frac{2\,N_c}{\as\,\pi^2}\,\,\int^{Q^2}_0\,d 
k^2\,\,\int\,d^2 b_t\,\,\tilde{N}(y,\xi,b_t)\,\,.
\eeq

The simplest way to understand \eq{GNRE} is to notice that at short 
distances the dipole amplitude in the coordinate space is directly 
related to the gluon structure function. Indeed, the dipole total cross 
section is equal to \cite{DIXS}
\beq \label{RESD}
\sigma_{dipole}(x,r^2_t)\,\,=\,\,\frac{\as(\frac{4}{r^2_t})}{N_c}\pi^2 
\,r^2_t \,xG(\frac{4}{r^2_t},x)\,\,\,=
\eeq
$$
\,\,\,2\,N(x, y; b_t) = 
2 \,r^2_t\,\int^{\infty}_0\,\,k\,dk \,\,\int \,d^2 b_t
\,J_0(kr_t)\,\tilde{N}(k,y;b_t)\,\,.
$$
The last relation is \eq{EQ1} in which we use a new notation for the 
dipole size ($r_t$).
Using moment representation for both $xG$ and $\tilde{N}$ in \eq{RESD}, 
namely
\beq \label{M}
xG(Q^2,x) \,\,=\,\,\frac{1}{2 \pi \,i}\,\int \,\,d \omega\,g(\omega)\,x^{- 
\omega}\,e^{\gamma(\omega)\,\ln Q^2}\,\,,
\eeq
we can find the relation of \eq{GNRE}  repeating simple calculation given 
in Ref.\cite{GLMVP}.

\begin{figure}[htbp]
\begin{minipage}{11.0cm}
\epsfig{file= 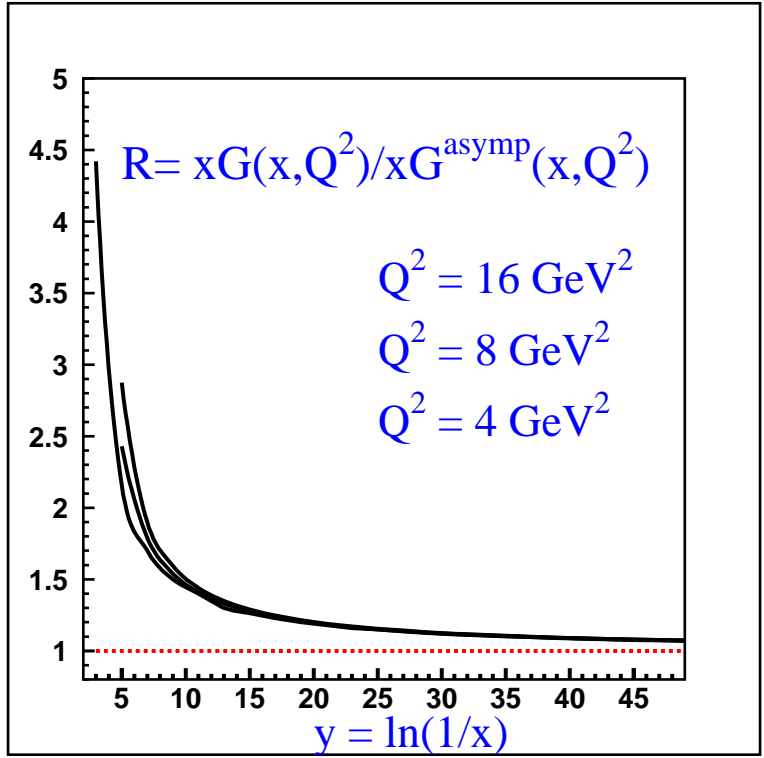,width=100mm}
\end{minipage}
\begin{minipage}{5.0 cm}
\caption{The ratio of the gluon structure function in the saturation (CGC) 
region  to it's 
asymptotic value given by \eq{GASYM} as a function of energy ($y = 
\ln(1/x)$) at different values of $Q^2$. }
\label{ratg2gas}
\end{minipage}
\end{figure}

Using asymptotic expression for $\tilde{N} = \h z$ (see \fig{scphi}) one 
can calculate $xG^{asymp}$ using \eq{GNRE}. We can take  all integrals 
analytically if we assume a simplified version of \eq{QSLB}, namely,
$Q^2_s(y; b_t) = Q^2_s(y; b_t=0)\,Exp(- 2m_\pi b_t)$ or
\beq \label{SIMZ}
 z \,\,=\,\,\bas\,\frac{\chi(\gamma_{cr})}{1\,-\,\gamma_{cr}}\,y 
\,\,-\,\,\ln k^2 \,\,-\,\,2\,m_\pi b_t \,\,-\,\,z_0\,\,=\,\,\ln 
(Q^2_s(y)/k^2) \,\,-\,\,2 \,m_\pi b_t\,\,-\,\,z_0\,\,.
\eeq
Integrating over $b_t$ in the saturation region ($z>0$ or 
$Q^2\,<\,Q^2_s(y,b_t=0)$)  we obtain
\beq \label{GASYM}
xG^{asymp}(Q^2,x)\,\,=\,\,\frac{2}{\pi\,\as}\,\frac{Q^2}{4 
m^2_\pi}\,\ln^3(Q^2_s(y)/Q^2)\,\,.
\eeq

\fig{ratg2gas} shows the ratio of $xG$ calculated directly from \eq{GNRE} 
using the numerical solution which we have discussed in the previous 
section to the asymptotic expression of \eq{GASYM}. Actually only to find 
out this difference we developed the method of solving for the master 
equation since the asymptotic behavior is clear from the general 
properties of the dipole scattering amplitude, namely, from the fact that 
$N(r^2_t,y,b_t) \,\leq\,1$ due to the unitarity constraint.

In \eq{GASYM} we consider so high energy that the upper limit in $b_t$ 
integration which is equal to $b_t=b_0 =\ \frac{1}{2\,m_\pi} 
\,\ln(Q^2_s(y)/k^2)$ is larger than $1/2 m_\pi$. In the region where 
$b_0\,<\,1/2 m_\pi$ we can use the pure pQCD expression for the saturation 
momentum given by \eq{QSPER}. The value of $b_0$ is equal to
\beq \label{B0P}
b^4_0\,\,=\,\,\frac{1}{k'^2 \,k^2}\,\pi\,\as^2
\frac{\,(N^2_c - 1)}{6\,N^2_c}\,\,\,e^{\as
\,\frac{\chi(\gamma_{cr}}{1 \,- \,\gamma_{cr}}\,( y - y_0)}\,\,.
\eeq

Using \eq{B0P} and \eq{GNRE} we have
\beq \label{GASYMPE}
xG^{asymp}(Q^2,x)\,\,=
\eeq
$$\frac{1}{\pi\,\as}\,\int^{Q^2}_0\,d\,k^2\,\,2
\int^{b^2_0}_0\,\,d\,b^2\,\ln(b^2_0/b^2)\,\,=\,\,4\sqrt{\frac{(N^2_c - 
1)}{6\,N^2_c\,\pi}}\,\frac{Q}{k'}\,e^{\as\,\h\,\,
\,\frac{\chi(\gamma_{cr}}{1 \,- \,\gamma_{cr}}\,( y - y_0)}\,\,. 
$$

Therefore, the gluon structure function shows a power-like increase for 
 $y$ smaller $y_a$ given by equation $b_0(y_a) = 1/2\,m_\pi$ with $b_0$ 
taken from \eq{B0P}.

For running $\as$ we have to resolve \eq{B0} to find $b_0(y)$ and 
substitute it to \eq{GASYMPE}. It is easy to see that in the pQCD region
\beq \label{B0RA}
b^2_0\,\,=\,\,\frac{1}{k\,k'}\,\,\,\bar{\tau}_s\,
e^{\,\omega_L\,(y_{in}
- y_0)}\,\,e^{\sqrt{\frac{8\,\pi}{b}
\chi'_{\gamma_{cr}}(\gamma_{cr})\,(y\,-\,y_0\,)}}.
\eeq
However, at $y=y_a$ $b_0$ from \eq{B0RA} reaches the value $b_0 = 
\frac{1}{2\,m_\pi}$ and for $y > y_a$ the value of $b_0$ is equal to
\beq \label{B0RA1} 	
b_0\,\,=\,\,\frac{1}{2\,m_\pi}\,\,\sqrt{\frac{8\,\pi}{b}
\chi'_{\gamma_{cr}}(\gamma_{cr})\,(y\,-\,y_0\,)}.
\eeq

$\mathbf{\sigma}${\bf(dipole)}.  This cross section is equal to
\beq \label{DIXS}
\sigma(dipole)\,\,=\,\,2\,\,\int \,d^2b_t\,N(r_t,y;b_t) \,\,.
\eeq
We use \eq{EQ1} to calculate $N(r_t,y;b_t)$ and it is easy to do for the 
asymptotic formula for $\tilde{N}(k,y;b_t) = \h z $ $  
\,=\,\ln(Q_s(y;b_t)/k)\,=\,K_0(k/Q_s(y,b_t))$. Here, we replace log 
contribution by McDonald function which reproduces this log at small value 
of the argument and leads to an exponential decrease at $k \,\geq\,Q_s$. 
For such large $k$ we cannot trust our simple formula and have to replace 
it by the perturbative QCD expression which will give a small contribution 
in the total cross section due to the fast decrease at large $b_t$.
The integral over $k$ in \eq{EQ1} give $N(r_t,y;b_t) = 1$.   
Integration over $b_t$ we can take using $z$ from \eq{SIMZ} with changing 
$k\, \rightarrow \,2/r_t$. Therefore the asymptotic behavior for 
$\sigma(dipole)$ is
\beq \label{DIXS1}
\sigma(dipole)\,\,=
\,\,\frac{2 \pi}{4\,m^2_\pi} \,\ln^2 (r^2_t\,Q^2_s(y,b_t=0))\,\,.
\eeq

For $y \,\leq \,y_a$ the upper limit in $b_t$ integration should be taken 
from \eq{B0P} and 
\beq \label{DIXS2}
\sigma(dipole)\,\,= 2 \pi b^2_0(y)\,\,=\,\,\h\,\pi\,\,r_1\,r_2 \,\,\as
\sqrt{\pi \frac{\,(N^2_c - 1)}{6\,N^2_c}}\,\,\,e^{\,\h\,\as
\,\frac{\chi(\gamma_{cr}}{1 \,- \,\gamma_{cr}}\,( y - y_0)}\,\,.
\eeq

{\bf $\mathbf{\gamma}^\mathbf{*} \mathbf{-} \gamma^\mathbf{*} $   cross 
section.}

The total cross section for $\gamma^* - \gamma^*$ interaction can be 
written as follows (see \fig{gaga}):
\beq \label{PHPHXS}
\sigma(Q_1,Q_2,W) \,\,=
\eeq
$$
\,\,\int^1_0 d z_1 \int^1_0 d z_2\,
\int \,d^2 r^2_{1,t} \,\int\,d^2 
r_{2,t}\,\,|\Psi(Q_1;z_1,r_{1,t})|^2\,|\Psi(Q_2;z_2,r_{2,t})|^2\,\,
\sigma(dipole)
$$
where $\sigma(dipole)$ is given by \eq{DIXS} while the wave functions of 
virtual photons are well known \cite{WF}.

The asymptotic expression using \eq{DIXS1} has been written and discussed 
in Ref. \cite{BKL}. Here, we can add to discussions in this paper the fact 
that the asymptotic behavior starts rather early as one can see from 
\fig{scphi}, namely, at $y\,\geq\,2 \div 2.5$.

All above estimates were done for the frozen QCD coupling because only in 
this case we obtained the semi-classical solution in entire kinematic 
region. For running $\as$ we have not solved the master equation since (i) 
we 
could not prove the geometrical scaling behavior in the saturation (CGC) 
domain; and (ii) we do not know what is a correct way to take into account 
the running QCD coupling for the BFKL emission. 
However, all asymptotic formulae have so transparent physical  
meaning that we can write them for running QCD coupling as well.

Indeed, \eq{GASYM} has three logs:
\begin{itemize}
\item \quad  One comes actually from the fact 
that the dipole amplitude in the coordinate space should approach 
unity 
at high energies. This is a direct consequence of the unitarity 
constraint and, because of this, will be the same for any solution to 
the master equation;

\item \quad Two logs are originated from the impact parameter 
integration. The limit for this integration stems from the kinematics 
of the saturation (CGC) domain, namely, from the fact that $k^2 
\,\leq\,Q^2_s (y,b_t)$ in this region. This relation leads to the 
$b_t \,<\,\frac{1}{2 m_\pi}\, \ln (Q^2_s/k^2)$ as we 
found in \eq{SATSCR}.
\end{itemize}

Therefore, we expect the same form of the asymptotic behaviors for 
running $\as$ as for frozen QCD coupling in terms of the saturation 
scale and the size of scattering dipoles. However, the energy 
behavior of the unitarity bounds will be quite different since the 
saturation scale is proportional to $ Q^2_s\,\propto 
\,\sqrt{\ln(1/x)} $ for the running QCD coupling while it increases as 
$\ln(1/x)$ for constant $\as$.

\section{Accuracy of the approach}

As we have discussed we neglected the contributions to the non-linear 
term in \eq{EQ} from the region of integration  $x_{12} \,\approx\,x_{02} 
\approx\,2 b_t\,\gg\,r_2$. To estimate the accuracy of our approach we 
have to calculate the term that we neglected using the solution to the 
non-linear equation that we found in this paper.

\begin{figure}
\begin{minipage}{11cm}
\epsfig{file=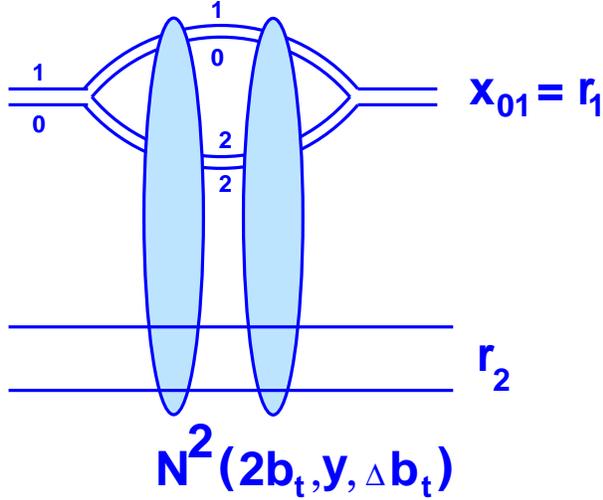,width=80mm}
\end{minipage}
\begin{minipage}{5cm} 
\caption{ Corrections to the non-linear equation due to interaction of 
two large dipoles ($x_{12}\, \approx\, x_{02}\,\approx\,2b_t 
\,\gg\,r_2$).}
 \label{cn2}
\end{minipage}
\end{figure}

We need to calculate the diagram of \fig{cn2} which is equal (see 
\eq{NLTERM1} and discussion around this equation )
\beq \label{CN21}
\as(x_{01})\,\int\,d^2 \Delta b_t \frac{x^2_{01}}{(b_t \,+\,\Delta 
b_t)^4}\,\,N^2(2( b_t + \Delta b_t),y,\Delta b_t)\,\,.
\eeq
 For 
$b_t\,\,\gg\,\,b_0(y)$ where $b_0(y)$ is the radius of interaction that 
stems from our solution, it is easy to see that \eq{CN21} leads to
the contribution which is of the order of
$$
\Delta N \,\approx\,\as(x_{01})\frac{x^2_{01}\,b^2_0(y)}{b^4_t} 
$$
since the dipole amplitude $N$ is of the order of unity for $b_t < b_0$.
One can see that $\Delta N \,\,\ll\,\,N$ and we can safely neglect this 
contribution.

For $r_2\,\ll\,b_t\,\,\ll\,\,b_0(y)$ \eq{CN21} gives
$$
\Delta N \,\approx\,\as(x_{01})\,\frac{x^2_{01}}{b^2_0(y)}\,\,.
$$
Since $x_{01} \,\ll\,b_0(y)$ and $b_0(y)$ increases at least 
logarithmically with energy $\Delta N$ is very small.

 To the right of the critical line the non-linear term of the evolution 
equation is small and the diagram of \fig{cn2} has an additional smallness 
which is proportional to $\as(x_{01})\,\frac{x^2_{01}\,r^2_2}{b^4_t}$ for 
$b_t \,>\,r_2$.

It is interesting to notice that we can repeat the same estimates for the linear term in 
\eq{EQ} replacing $N^2$ in \eq{CN21} by $2N$. These estimates  would lead to the same 
result. It 
means that the large $b_t$ dependence of the linear term (see, for example,  \eq{QSPER} 
) stems from the dipole sizes smaller than the size of the target. This is a transparent 
explanation why the first linear term in \eq{EQ} has the form given by \eq{EQK} 
without any additional assumption about the size of the dipoles in the non-linear 
equation \footnote{We thank our referee who pointed out this transparent illustration 
of the form of \eq{EQK}.}.

The most important restriction of the applicability of \eq{EQ} stems from 
enhanced diagram of \fig{fan}-c. Estimating these enhanced diagrams we should 
distinguish  two different kinematic  regions : $ k^2 > Q^2_s$ and $k^2 < Q^2_s$. 
In the first one the integration over $y_2$ in \fig{fan}-b converges at $y_2 \propto 
1/\bas\,\,\ll\,\,y$ and, therefore, the enhanced diagram reduces to the diagrams of 
\fig{fan}-a ( see Refs. \cite{GLR,MUQI,SAT,ELTHEORY,KV} for details). In this region the 
nuclear target has an advantage that the `fan' diagram originated from the enhanced 
diagram has an additional suppression with respect to parameter $\as A^{\frac{1}{3}}$ 
\cite{KV,SCH}.

However, the most interesting region is the second one ($k^2 < Q^2_s$). In this region 
there is no reason to expect that the integral over $y_2$ will be concentrated at $y_2 
\ll y$ and the enhanced diagrams will lead to the dominant contribution. The nuclear 
target has no advantages  in this region since the solution here does not depend on the 
initial 
conditions. As one can see directly from \eq{EQ} the particular properties of  target 
enters only in the initial conditions.  We 
are concentrated our efforts on dealing  with this kinematic region in this section.

The expression for the first correction due to enhanced diagram to  ``fan"  
diagrams presented in \fig{fan}-c is  
the following \cite{GLR,KV}:
\beq \label{ENDI}
\int d^2 b_3 \Delta N^{enhance}\,\,=
\,\,\frac{\bas^2}{N^2_c}\,\,\int 
\,d^2b_{3} 
\int^y_0 
\,d y_1 \int\, d^2 
k_1\, d^2 b_{1}\int^{y_1}_0 \,d y_2\,\int\, d^2 k_2 d^2 b_{2}
\eeq
$$
\tilde{N}(k,k_1,y-y_1,b_{1}) \tilde{N}^2(k_1,k_2,y_1-y_2,b_{2}) 
\tilde{N}(k_1,k',y_2,b_{3})
$$

All notations are clear from \fig{fan}-b and \fig{fan}-c.
It should be stressed that all amplitudes in \eq{ENDI} except the last 
one($\tilde{N}(k_1,k',y_2,b_{3}) $) are different from the solution of the 
non-linear equation (see \fig{fan}-c). They actually satisfy a different 
equation which it is easy to write. Indeed, the slight glance at 
\fig{fan}-c shows that the equation for $\tilde{N}(k,k_1,y-y_1,b_{1}) $ 
can be written in the form:
\beq \label{EQNEW}
 \frac{\partial \tilde{N}(k,k_1,y - y_1;b_t)}{ \partial
\,y}\,\,=
\eeq
$$
\,\,\bas\,\left(\,\hat{\chi}(\hat{\gamma}(k))\tilde{N}(k,k_1,y-y_1;b_t)\,
\,-
\tilde{N}(k,k_1,y-y_1;b_t)\,\tilde{N}(k,k',y;b_t)\right)\,\,;
$$
where $\tilde{N}(k,k',y;b_t)$ is a solution of \eq{EQK}. Since the 
solution to this equation is a function of one variable $z$ and at $y_1=0$ 
it  should coincide with the solution to \eq{EQK} one can find that 
$\tilde{N}(k,k_1,y-y_1;b_t)\,\,=\,\,\tilde{N}(z - z_1)$ while 
$\tilde{N}(k,k',y;b_t) = \tilde{N}(z)$  where $N(z) $ is 
defined in \eq{SRSOL1}. We can rewrite the argument of 
$\tilde{N}(k,k_1,y-y_1;b_t)$ in more convenient form using \eq{B0} and 
\eq{B0RA}, namely
\beq \label{N2}
\tilde{N}(k,k_1,y-y_1;b_t)\,\,=\,\,\tilde{N}\left(\frac{F(y 
-y_1)\,S(b_t)}{k^2\,k^2_1\,b^4_t}\right)
\eeq
where $F(y -y_1)$ is a function of rapidity. This function is different 
for frozen and running $\as$, namely $F(y) \,\propto \,e^{\lambda_r\,y}$ 
for constant $\as$ and $F(y) \,\propto \,e^{\lambda_r\sqrt{\,y}}$ for 
running QCD coupling (see \eq{B0} and \eq{B0RA}).
$S(b_t)$ is a non-perturbative profile function for which we assume that 
$S(b_t)\,\,\rightarrow\,\,e^{- 2 m_\pi b_t}$ at large $b_t$ 
($b_t\,\geq\,1/2\,m_\pi$).
 Using \eq{N2} we can 
evaluate  all integrals in \eq{ENDI}. One can see that in pQCD region for 
not very high energies where we can consider that $S(b_t) = 1$ we have  
\beq \label{N21}
\int d^2 b_3 \Delta 
N^{enhance}\,\,\propto\,\,\frac{\bas^2}{2\,N^2_c}\,
\frac{\ln^2(k^2/k'^2)}{k\,k'}\,\,
\int^y_0 dy_1 \int^{y_1}_0 d y_2 F(y - y_1)\,F(y_1 - y_2) \,F(y_2)\,
\eeq

Calculating the ration of $\Delta N$ to $N$ to characterize the scale of 
corrections we obtain:
\beq \label{RATIO}
\frac{ \int d^2 b_3 \Delta N}{\int d^2 b_t 
N}\,\,\propto\,\,\frac{\bas^2}{N^2_c}\ \frac{\int^y_0 dy_1 \int^{y_1}_0 d 
y_2 F(y - y_1)\,F(y_1 - y_2) \,F(y_2)}{F(y)}\,\,.
\eeq

For frozen $\as$ this ratio is of the order of $\frac{\bas}{N^2_c} y^2$
\footnote{We do not consider $\ln(k^2/k'^2)$ here 
as a large factor, since 
in the saturation region this factor should not be specially large.} 
which means that we can neglect the enhanced diagrams only in limited 
range of energies. However for running $\as$ this ratio is proportional to
$\frac{\bas}{N^2_c} e^{\lambda_r (\sqrt{3} - 1)\sqrt{y}}$ and the range of 
energy (rapidity) where we can still neglect the enhanced diagrams shrinks 
even more than for constant $\as$.

At first sight  the above discussion is in the direct contradiction with the paper 
of Mueller and Patel \cite{MUPA} where was shown that (i) the enhanced diagram  of 
\fig{fan}-b has no extra $N_c$ suppression with respect to  `fan' diagrams of 
\fig{fan}-a; and (ii) the normalization of $N$ could be chosen in a such way that extra 
factor $1/N^2_c$ will not appear in the calculation.

To clarify our calculations we will repeat them in a toy-model: `fan'  diagrams for 
Pomeron exchanges with $\alpha'_P =0$ and intercept $\alpha_P(0)= 1 + \Delta$. For the 
nuclear target such model correspond to 
Schwimmer resumation \cite{SCH} and describes the hadron - nucleus interaction at high 
energy. In this model the dipole-dipole interaction is given by sum of `fan' diagrams   
of \fig{fan}-a, namely,
\beq \label{TM1}
\tilde{N}(y) \,\,\,=\,\,\,\frac{g_1\,g_2 e^{\Delta y} 
}{1\,+\,g_2\,\frac{G_{3P}}{\Delta}\,\left( 
\,e^{\Delta 
y}\,-\,1\right)}\,\,\,\,\,\,\stackrel{y \gg 1}{\longrightarrow}\,\,\,\,\,\,\,\frac{g_1 
\Delta}{G_{3p}}\,\,\,\,\,\,\,=\,\,\,\,\,\,\,1
\eeq 
where $g_1$ and $g_2$ are vertices of Pomeron interaction with dipoles $1$ and $2$ ($r_1 
\,\ll\,r_2$).Actually, $g_2$ contains an extra factor $1/N^2_c$ as one can see from 
\eq{INPHI}. In \eq{TM1} we assumed that $N(y) \,\rightarrow\,1$ at $y \,\rightarrow 
\infty $ as it follows from \eq{EQ}. The subset of `fan' diagrams that starts from the 
Pomeron and finishes at dipole  $2$ (target) leads to the answer
\beq \label{TM2}
\tilde{N}(y' - 0)\,\,\,\,\,\,\,=\,\,\,\,\,\,\frac{G_{3P}\,g_2 e^{\Delta (y' -0)} 
}{1\,+\,g_2\,\frac{G_{3P}}{\Delta}\,\left(
\,e^{\Delta
y}\,-\,1\right)}\,\,\,\,\,\,\stackrel{y \gg 1}{\longrightarrow}\,\,\,\,\,\,\,\,\Delta 
\Theta(y'-0)\,\,,
\eeq
where $\Theta  (y) $ is equal to 1 for $y > 0$ being zero for negative $y$.
 
One can see that $\tilde{N}(y_1 - y_2)$ starts with one Pomeron exchange $ 
e^{\Delta(y_1 - y_2)}$. Summing all `fan' diagrams attached to this Pomeron we have
\beq \label{TM3}
\tilde{N}(y_1 - y_2)\,\,=\,\,\,e^{\Delta(y_1 - y_2)}\,\sum_{n=0}\,\frac{(-1)^n}{n!}\,
\left( \Delta\,(y_1 - y_2) \right)^n\,\,=\,\,1\,\,;
\eeq
while
\beq \label{TM4}
\tilde{N}(y - y_1)\,\,=\,\,g_1\,e^{\Delta(y - y_1)}\,\sum_{n=0}\,\frac{(-1)^n}{n!}\,
\left( \Delta\,(y - y_1) \right)^n\,\,=\,\,g_1\,\,;
\eeq
and
\beq \label{TM5}
\tilde{N}(y_2 - 0)\,\,=\,\,\frac{\,g_2 e^{\Delta y_2} 
}{1\,+\,g_2\,\frac{G_{3P}}{\Delta}\,\left(
\,e^{\Delta
y_2}\,-\,1\right)}\,\,\,\,\,\,\stackrel{y \gg 1}{\longrightarrow}\,\,\,\,\,\,\,\frac{
\Delta}{G_{3p}}\,\,.
\eeq

Collecting \eq{TM2} - \eq{TM5} we obtain for \eq{ENDI} and taking into account that
$\frac{g_1\Delta}{G_{3p}}\,\,=\,\,1$ we obtain
\beq \label{TM6}
\Delta N^{enhance}\,\,=\,\,\frac{G_{3P}}{N^2_c} \int d y_1 \int dy_2 \,\propto
\frac{\bas^2}{2\,N^2_c}\,y^2
\eeq
which coincide  with \eq{RATIO}.

Formally speaking the enhanced diagrams is suppressed by factor $1/N^2_c$ 
and have not been included in the non-linear equation as well as a large 
number of other diagrams of the order of $1/N^2_c$. Therefore, our 
estimates show that $1/N_c$ corrections could be rather large. Actually, 
we know how to calculate such kind of corrections ( see paper of Balitsky 
in Ref. \cite{KV} and   paper of Wiegert in 
Ref.\cite{ELTHEORY})\footnote{We thank Heribert Weigert for discussion of 
$1/N_c$ corrections. He shared with us his first attempts to calculate 
numerically these corrections.}

\section{Discussion and Conclusions}

In this paper we discussed the semi-classical approach to the solution of 
the non-linear equation which has been suggested and developed before 
\cite{GLR,SCA,SAT,LT}. The new domain which we explored in the 
semi-classical approach is the saturation 
(CGC) region which was out of reach by this method before. The new result 
is also related to the impact parameter dependence which has not been 
considered in Refs.\cite{GLR,SCA,SAT} and has been approach only in 
the simplified case of double log approximation of pQCD in Ref. \cite{LT}.

We hope that we demonstrated the main advantages of the semi-classical  
approach, namely
\begin{itemize}
\item \quad The natural appearance of the new saturation scale as the
momentum of the critical line which divides the set of trajectories in two 
parts: the first group is  trajectories which approach the trajectories 
of 
the linear equation at large value of $\xi = \ln(k^2k'^2b^4_t) $ (see 
\fig{gaga} for notations); and the second group   is trajectories that go 
apart from the trajectories of the linear equation;
\item \quad The solution in the saturation (CGC) domain reproduces the 
geometrical scaling behavior , namely, $\sigma_{dipole}(r_t,x;b_t)$ turns 
out to be a 
function of one variable $\tau = r^2_t\,Q^2_s(x;b_t)$; 
\item \quad The impact parameter dependence of the saturation scale stems 
from the initial condition (Born approximation) as has been discussed in 
Refs. \cite{LRREV,FIIM} but not from the kernel of the linear evolution 
equation (BFKL kernel)) as was suggested in Ref.\cite{KW}. The large $b_t$ 
behavior of the saturation scale should be determined by non-perturbative 
contribution to the Born amplitude which  has  been considered in Ref. 
\cite{BKL}.
\end{itemize}
We found asymptotic behavior both for the dipole amplitude 
 $N$ and function $\Phi$ (see \eq{PHI}). It should be stressed that the 
behavior of function $\Phi$ at the large values of the argument $z$  is 
quite different from the double log approximation case discussed in 
Ref.\cite{LT}. It should be stressed that we found that the exact solution 
could  be approached by asymptotic one starting with 
$y=\ln(1/x) \,\approx\,2 \div 2.5$.

We also found the unitarity bound for the gluon structure function 
and for the dipole cross section, given by \eq{GASYM} and \eq{DIXS1}, 
which depends on the energy dependence of the saturation scale. The fact 
that high energy behavior of the dipole cross section depends on 
$\ln\left(Q^2_s(x;b_t=0)\,r^2_t \right)$ differs our analysis from the 
considerations discussed before \cite{LRREV,FIIM,KW}. 
It should be stressed that our formulae lead to quite different dependence 
on energy for the high energy asymptotic behavior.
Even for the frozen QCD coupling our results are different from the 
estimates in Refs. \cite{LRREV,FIIM}. For example, we predict that
\beq \label{CO1}
\sigma(dipole) \,\,\leq\,\,\frac{2 \pi}{4 \,m^2_\pi}\,\left( \as 
\frac{\chi(\gamma_{cr})}{1 - \gamma_{cr}}\,\ln(1/x) \right)^2
\eeq
 instead of
\beq \label{CO2} 
\sigma(dipole) \,\,\leq\,\,\frac{2 \pi}{4 \,m^2_\pi}\,\left( \as
\chi(\h)\,\ln(1/x) \right)^2
\eeq
 as was obtained in Refs.\cite{LRREV,FIIM}. It is  worthwhile mentioning that $\gamma_{cr} 
\approx 0.37$ (see \eq{GAMMACR}) is not related to the running QCD coupling. It stems 
from the full BFKL kernel while $\gamma_{cr} = \h$ arises only if we approximate the 
full BFKL kernel by the double log formula $\chi(\gamma) = 1/\gamma$.

In the case of running $\as$ the difference is even more remarkable since 
our formulae lead to a different energy behavior. Indeed, we have
\beq \label{CO3} 
\sigma(dipole) \,\,\leq\,\,\frac{2 \pi}{16 \,m^2_\pi}\,\left( 
\frac{4\,\pi}{b}\,\,
\frac{\chi(\gamma_{cr})}{1 - \gamma_{cr}}\,\ln(1/x) \right)
\eeq
as one can see from \eq{SATSCR}. Notice that extra factor 4 in the 
dominator stems from the different initial conditions that we used in 
the case of running $\as$.  

\begin{figure}
\begin{minipage}{10.0cm}
\epsfig{file= 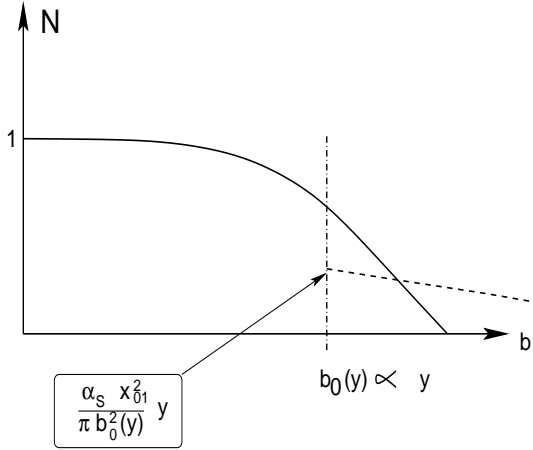,width=70mm,height=60mm }
\end{minipage}
\begin{minipage}{6.0 cm}
\caption{The dipole amplitude as a function of the impact parameter $b$ as it emerges 
from discussion in section 6 (see \eq{CN21} ). The dotted line shows the correction 
term evaluated using \eq{CN21} .} \label{nprof}
\end{minipage}
\end{figure}

Actually, the discussion about corrections to the non-linear master equation given in 
section 6 can be summarize in the picture shown in \fig{nprof}. The corrections from the 
dipole which sizes larger than the size of the target lead to a large $b$ tail but it 
starts at very low value and it could not contribute to the energy behaviour of the 
total cross section. The entire picture is the same  in the main features as it was 
suggested    in Ref. \cite{FIIM} 

We started this paper having in mind two questions to answer. The 
practical one: to find the analytical solution which everyone can check 
and which includes the impact parameter dependence. We firmly believe that 
only after finding such a solution we could develop the numerical methods 
of 
solving our master non-linear equation (see \eq{EQ}). Since the 
semi-classical approach is valid for the large values of $y = 
\ln(1/x)\,\gg\,\ln Q^2$ we can use this solution as a guiding one for the 
numerical attempts to solve the non-linear equation.

Our second motivation was to reach a theoretical understand how to 
incorporate the non-perturbative QCD  corrections in the nonlinear 
equation. These corrections have to be essential at large values of the 
impact parameters where they will change the power-like decrease in pQCD 
to the exponential one as follows from the spectrum of the observed 
hadron \cite{FROI}. We believe that we answered this question  by 
demonstrating that we should take the non-perturbative correction only in 
initial conditions (in Born amplitude). How to take into account these 
non-perturbative corrections in the Born amplitude was considered in our 
previous publications \cite{BKL}.

How to incorporate the non-perturbative corrections in initial conditions 
or in the kernel actually crucially depends on the strategy of the 
approach. If you consider first the solution of the linear equation you 
have to include the non-perturbative corrections at large $b_t$ in the 
BFKL kernel as was discussed in Ref. \cite{KW}. We believe that the 
correct strategy is to solve the non-linear equation. As has been 
discussed for long time \cite{GLR,LRREV,BAR,GMS} the non-linear dynamics 
leads to a suppression at low $x$  of the density of partons with   low 
transverse 
momenta ($p_t$). Since  due to the uncertainty principle $\Delta b_t p_t 
\approx 1$ only partons in the beginning of evolution (at $x \,\approx 
\,1$) could have large values of $b_t$. In other words only parton 
distribution in initial parton cascade is relevant for the large $b_t$ 
behavior.

\begin{figure}
\begin{minipage}{10.0cm}
\epsfig{file= 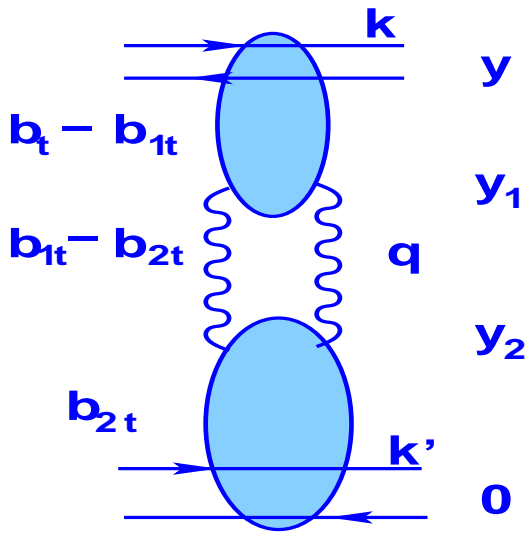,width=70mm,height=70mm }
\end{minipage}
\begin{minipage}{6.0 cm}
\caption{The diagram which describes the large $b_t$ behavior of the 
dipole amplitude.Two gluons are close to the mass shell.}
\label{deltak}
\end{minipage}
\end{figure}

The above statement sounds  strange because we claim that in the BFKL 
ladder we need to take into account the non-perturbative QCD correction 
only for two $t$-channel gluons attached to the larger dipole. Indeed, the 
large $b_t$ behavior stems from the exchange the lightest particles 
\cite{FROI}( 
gluon in our case) somewhere in the middle of the `ladder' (see 
\fig{deltak} ). Since in the middle of the ladder we have massless gluons 
the large $b_t$ behavior is determined by the diagram of \fig{deltak}. It 
has a form
\beq \label{LBB}
\tilde{N}(k,k',y;b_t)\,\,=
\eeq
$$\,\,\frac{1}{b^2_t}\,\int d^2 q \int^y_0\,d y_1\,\int d^2 
(b_t - b_{1t}) 
\tilde{N}(k,q,y - y_1;b_t - b_{1t})\,\,\int^{y_1}_0\,d y_2\ \int d^2 (b_t 
- 
b_{1t})
\tilde{N}(k,q, y_2;b_{2t})\,
$$
where we consider the $b_t$ is much larger than typical impact parameters 
in the integration over $(b_t - b_{1t})$ and $b_{2t}$.
Using \eq{N2} we can take integrals over impact parameters in \eq{LBB}. 
For the upper amplitude we have to consider $S(b_t) = 1$ in \eq{N2} since 
we do not introduce non-perturbative corrections, while for the lower 
amplitude we substitute our solution at large $b_t$ for which $S(b_t) 
\,\rightarrow\,e^{- 2 m_\pi b_t}$. Finally, we have
\beq \label{LBB1}
\tilde{N}(k,k',y;b_t)\,\,\rightarrow\,\,\frac{1}{b^2_t\,(2 
m_\pi)^2}\,\,\int^y_0 d y_1 
F(y - y_1)\,\int^{y_1}_0 d y_2 \,y^2_2.
\eeq
Since $F(y - y_1)$ behaves as an exponent the integral chooses small 
$y_2$. Therefore, actually the cell with large $b_t$ is the closest one to 
the largest dipole. This simple estimates show that our statement that the 
large $b_t$ behavior of the solution to the non-linear equation is 
determined by the initial condition is self-consistent .

\begin{figure}
\begin{minipage}{11.5cm}
\epsfig{file= 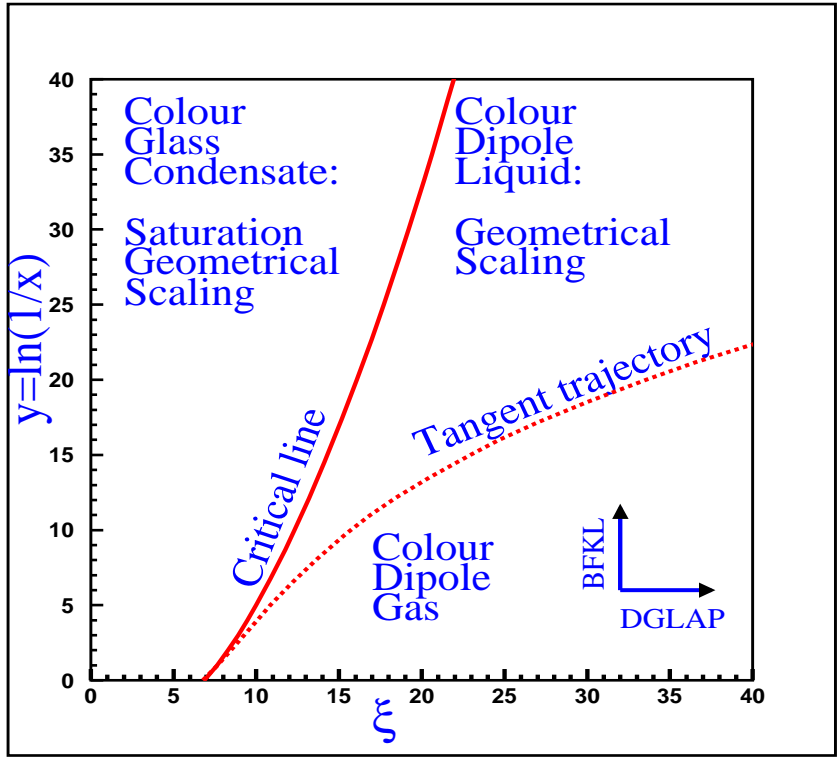,width=110mm,height=100mm}
\end{minipage}
\begin{minipage}{4.5 cm}
\caption{The phase map of high parton density QCD.}
\label{phase}
\end{minipage}
\end{figure}

The great advantage of the semi-classical approach is that we clearly 
see all phases of high parton density QCD (see \fig{phase}).
The critical line separates the phase with high parton densities.
To the left of critical line we have a phase  in 
which we have both saturation and the geometrical scaling behavior for 
the gluon density. This phase L.McLerran and his team suggested to call
``Color Glass Condensate" (CGC)  since the condensed system of color 
dipoles 
behaves as glass : a disordered system which evolves very slowly relative 
to a natural time scale (see more and better in a new review of this 
approach written by  Iancu and Venogapalan in Ref. \cite{SAT}).To the 
right of the critical line the parton system is still rather dense but it 
is far away from 
saturation while still has a geometrical scaling behavior \cite{IIM}.
This system be can call Color Dipole Liquid (CDL). 
The order parameter for $CGC \rightarrow CDL$ transition is the saturation 
scale given by the critical line.
The trajectory which is tangent to the critical line, separates the CDL 
phase from the phase with low density of color dipoles for which we can 
use the linear evolution equations: the BFKL or DGLAP ones. The natural 
name for this phase is Color Dipole Gas.

\section{Acknowledgments}
 
  We thank  M. Braun, E. Gotsman, U. Maor and A. Mueller for fruitful 
discussions on the subject 
of this paper.
  E.L.  is indebted to the Alexander-von-Humboldt Foundation for the 
award that gave him a possibility to work on low $x$ physics during the 
last year.

 This research was supported in part by the
GIF grant \# I-620-22.14/1999, and by the Israel Science Foundation,
founded by the Israeli Academy of Science and Humanities.

\end{document}